\documentclass[12pt,preprint,rotate]{aastex}
\usepackage{times}
\usepackage{graphicx}
\usepackage{natbib}
\usepackage{epstopdf}
\newcommand{\ltsima}{\stackrel{\textstyle <}{\sim}}
\newcommand{\simlt}{\scriptsize{\raisebox{-2pt}{$\ltsima$}}\normalsize}
\newcommand{\gtsima}{\stackrel{\textstyle >}{\sim}}
\newcommand{\simgt}{\scriptsize{\raisebox{-2pt}{$\gtsima$}}\normalsize}

\renewcommand{\thepage}{\rm\arabic{page}}

\newcommand{\swas}{{\em SWAS}}

\newcommand{\water}{H$_2$O}
\newcommand{\iwater}{H$_2^{\:18}$O}
\newcommand{\nwater}{H$_2^{\:16}$O}
\newcommand{\oxy}{O$_2$}
\newcommand{\nco}{$^{12}$CO}
\newcommand{\ico}{$^{13}$CO}
\newcommand{\cio}{C$^{18}$O}

\newcommand{\xwater}{{\it x}(o--H$_2$O)}

\newcommand{\asec}{$^{\prime\prime}$}
\newcommand{\amin}{$^{\prime}$}
\newcommand{\ddeg}{$^{\rm o}$}
\newcommand{\mh}{H$_2$}
\newcommand{\nh}{$n${(H$_2$)}}

\newcommand{\av}{{\em A}$_{\rm V}$}
\newcommand{\um}{$\:\mu$m}
\newcommand{\kms}{~km~s$^{{-1}}$}
\newcommand{\cmc}{cm$^{-3}$}
\newcommand{\cms}{cm$^{-2}$}

\newcommand{\vlsr}{$v_{\rm LSR}$}
\newcommand{\ta}{$T_{A}^{\;\;*}$}
\newcommand{\go}{$G_{\rm o}$}
\newcommand{\dash}{$\,$--$\,$}

\newcommand{\etal}{et$\,$al.}
\newcommand{\ti}{$\,\times\,$}

\renewcommand{\apj}{{\em Ap.~J.}}
\renewcommand{\apjl}{{\em Ap.~J.~Letters}}
\renewcommand{\apjs}{{\em Ap.~J.~Suppl.}}

\renewcommand{\mnras}{{\em M.N.R.A.S.}}

\renewcommand{\aap}{{\em A\&A}}
\renewcommand{\aj}{{\em A.~J.}}

\makeatletter
\def\fnum@figure{{Fig.~\thefigure}}
\long\def\@makecaption#1#2{
 \vskip 10pt 
 \setbox\@tempboxa\hbox{#1. #2}
 \ifdim \wd\@tempboxa >\hsize \unhbox\@tempboxa\par \else \hbox
to\hsize{\hfil\box\@tempboxa\hfil} 
 \fi}

\makeatother

\makeatletter
\def\ps@myheadings{\let\@mkboth\@gobbletwo
\def\@oddhead{\hbox{}\sl\rightmark \hfil \thepage}%
\def\@oddfoot{}\def\@evenhead{\thepage\hfil\sl\leftmark\hbox {}}%
\def\@evenfoot{}\def\sectionmark##1{}\def\subsectionmark##1{}}
\makeatother

\makeatletter
\def\subsection{\@startsection{subsection}{2}{\z@}{-3.25ex plus -1ex minus 
   -.2ex}{1.5ex plus .2ex}{ \em \rm}}
\makeatother

\newcounter{ncount}

\shorttitle{Distribution of Water Vapor in Molecular Clouds}
\shortauthors{Melnick et al.}

\begin{document}

%\phantom{0}

%\vspace{2.4in}

%\begin{center}
%{\large\bf Version 4}\\*[9mm]
%\today
%\end{center}

%\clearpage

%\phantom{0}

\renewcommand{\baselinestretch}{1.20}

\vspace{-0.6in}

\title{\bf Distribution of Water Vapor in Molecular Clouds}
\author{Gary J. Melnick, Volker Tolls} 
\affil{Harvard-Smithsonian Center for Astrophysics, 60 Garden Street, 
Cambridge, MA 02138}
\email{gmelnick@cfa.harvard.edu, vtolls@cfa.harvard.edu}
\author{Ronald L. Snell}
\affil{Department of Astronomy, University of Massachusetts, Amherst, MA 01003}
\email{snell@astro.umass.edu}
\author{Edwin A. Bergin}
\affil{Department of Astronomy, University of Michigan, 825 Dennison 
Building, Ann Arbor, MI 48109}
\email{ebergin@umich.edu}
\author{David J. Hollenbach}
\affil{SETI Institute, 515 North Whisman Road, Mountain View, CA 94043}
\email{dhollenbach@seti.org}
\author{Michael J. Kaufman}
\affil{Department of Physics and Astronomy, San Jose State University, 
One Washington Square, San Jose, CA 95192-0106}
\email{mkaufman@email.sjsu.edu}
\author{Di Li}
\affil{Jet Propulsion Laboratory, 4800 Oak Grove Drive, Pasadena, CA 91109}
\email{dili@jpl.nasa.gov}
\and
\author{David A. Neufeld}
\affil{Department of Physics and Astronomy, Johns Hopkins University,
3400 North Charles Street, Baltimore, MD 21218}
\email{neufeld@pha.jhu.edu}

\begin{abstract}
We report the results of a large-area study of water vapor along the Orion Molecular Cloud ridge,  the
purpose of which was to determine the depth-dependent distribution of gas-phase water
in dense molecular clouds.  We find that the water vapor measured toward 77 spatial positions
along the face-on Orion ridge, excluding positions surrounding the outflow associated with
BN/KL and IRc2, display integrated intensities that correlate strongly with known cloud surface
tracers such as CN, C$_2$H, \ico\ $J =\:$5\dash 4, and HCN, and less well with the
volume tracer N$_2$H$^+$.  Moreover, at total column densities corresponding 
to \av$\:<\:$15 mag., the ratio of \water\ to \cio\ integrated intensities 
shows a clear rise approaching the cloud surface.  We show that this behavior cannot be accounted
for by either optical depth or excitation effects, but suggests that gas-phase water abundances 
fall at large \av.  These results are important as they affect 
measures of the true water-vapor abundance in molecular clouds by highlighting the 
limitations of comparing measured water vapor column densities with such traditional
cloud tracers as \ico\ or \cio.  These results also support cloud models that incorporate
freeze-out of molecules as a critical component in determining the depth-dependent abundance
of water vapor.
\end{abstract}


\vspace{4.5in}

\baselineskip=22pt

\clearpage

\baselineskip=22pt
\setcounter{page}{1}

\section{\bf Introduction}

\vspace{-1.1mm}

Interstellar water is of continuing interest because of the
role water plays in the oxygen chemistry within dense molecular clouds
as well as the efforts to trace its abundance and distribution
in all phases of cloud evolution through to planet formation.  
Thanks to a number of space-based observatories operated during the past
15 years, good progress has been made detecting and mapping the 
distribution of water toward molecular clouds.  The highest water 
abundances, and the strongest water emission, 
are observed toward warm (i.e., $T>\:$300~K) gas regions, most
frequently associated with shock-heated gas generated by high-velocity
outflows from young stellar objects and supernovae remnants.  This finding
is in agreement both with predictions that neutral-neutral reactions (i.e.,
H$_2$ + O $\rightarrow$ OH + H and  H$_2$ + OH $\rightarrow$ H$_2$O + H)
dominate at these temperatures and are relatively efficient at producing water
(cf.~Elitzer \& de Jong 1978; Elitzur \& Watson 1978), and that non-dissociative shocks may be
effective at liberating and heating water from ice mantles (Melnick~\etal\ 2008).  
Observations suggest that typically between 
1 and 20 percent of the elemental oxygen is driven into gas-phase \water\ via these
processes (cf.~Harwit~\etal\ 1998; Melnick~\etal\ 2000; Neufeld~\etal\ 2000b;
Nisini~\etal\ 2000; Benedettini~\etal\ 2002; Franklin~\etal\ 2008).

\pagestyle{myheadings}
\markright{{{}}\hfill{\hbox{\hss\rm Page \rm }}}

However, within most dense (\nh\ $>\,$10$^3$~\cmc) molecular cloud complexes, 
the bulk of the water -- vapor plus ice -- lies within the cooler ($T\,\simlt\,$40~K) 
and more massive quiescent gas
component.  Knowledge of the depth-dependent abundance of water vapor and
water ice in molecular clouds is important for at least two reasons.  First, the depth-dependent
abundance of water vapor and water ice reflects a competition among a number of 
important processes, such as photodissociation, photodesorption, 
gas-phase reactions, gas-grain reactions, and grain-surface reactions, most of which 
depend upon the gas density and 
far-ultraviolet flux (FUV; 6 eV$\,<\!h\nu\!<\,$13.6 eV).
A better understanding of these processes
and their relative importance thus reduces the uncertainty in virtually all models of the
chemical composition of molecular clouds.  Second, because oxygen is the most abundant
element after hydrogen and helium, the processes that control the amount of oxygen 
locked in water vapor and, in particular, water ice determine the amount of residual 
oxygen free to react with other species.  In this way, the predicted abundance of a 
host of species that depend on the gas-phase oxygen-to-carbon or oxygen-to-nitrogen 
ratio, for example, hinges on knowledge of the main reservoirs of oxygen, such as
gas-phase water and water-ice.

Models incorporating the formation and destruction processes mentioned above
have been constructed and detailed predictions exist for the water-vapor
and water-ice abundance profiles as functions of cloud density and external
FUV flux (cf.~Hollenbach~\etal\ 2009).
Measures of the strength of
%various different 
solid-state \water\ absorption features along numerous lines of sight
%with varying dust extinction 
provide good column density distributions for water ice 
(e.g.,~Whittet~\etal\ 1998; Sonnentrucker~\etal\ 2008).
Unfortunately, complementary studies of the distribution of gas-phase \water\ 
within quiescent molecular gas have suffered either from a lack of access to
the ground-state ortho- and para-water transitions, which probe most of the water
column at $T <\,$40~K, or from sparse spatial sampling of most
clouds.  In this paper we report the results from a large-area, fully-sampled study of 
ground-state water vapor emission
toward the Orion Molecular Cloud ridge using the {\em Submillimeter Wave Astronomy 
Satellite} ({\em SWAS}).

The {\em SWAS} mission was
primarily dedicated to the study of: (1) the oxygen chemistry
in dense (\nh$\,\simgt\,$10$^3$~\cmc) molecular clouds throughout our Galaxy;
(2) the abundance, distribution, and cooling power of \water\ and \oxy\ within
these clouds; and, (3) the structure and physical conditions in molecular clouds.
To achieve these goals, \swas\ was designed to detect emission
from five key gas-phase atoms and molecules -- water (H$_2^{\;16}$O)\footnote[1]
{$\,$Henceforth, the most abundant isotopologue of water, H$_2^{\;16}$O, will
be denoted simply as \water.}, isotopic water (\iwater), molecular oxygen 
(\oxy), atomic carbon (CI), and isotopic carbon monoxide (\ico).  
Since the emphasis was on studying the bulk of the colder ($T<\,$40~K)
molecular material, \swas\ measured those frequencies coinciding with
either the ground-state or a low-lying transition in each of these species.  
The one exception was \ico,
for which the mid-level ($E_{upper}/k\,\simeq\,$79~K) $J\,=\,$5$\,-\,$4 transition 
was observed.  Table~1 presents a summary of the species and transitions 
observed by \swas.  A detailed description of the {\em SWAS} mission can
be found in Melnick~\etal\ (2000b) and Tolls~\etal\ (2004).

The Orion Molecular Cloud ridge is an approximately 10\dash 15 arcminute-wide region
of warm ($T\,\sim\,$20$\,$--$\,$40~K), dense (\nh$\:\simgt\:$10$^4\,$--$\,$10$^6$~\cmc) gas 
(cf.~Bally~\etal\ 1987; Dutrey~\etal\ 1991;
Tatematsu~\etal\ 1993;
Bergin~\etal\ 1994; Bergin, Snell, \& Goldsmith 1996;
Ungerechts \etal\ 1997; Johnstone \& Bally 1999)
stretching $\sim\,$30 
arcminutes north and more than 60 arcminutes south of BN/KL.  As such, the
Orion ridge represents the contiguous region with the largest angular size observed
in water vapor
by {\em SWAS} and provided the opportunity to obtain 86 independent spatial
samples with {\em SWAS}'s 3.3\ti 4.5 arcminute beam (at 557~GHz).  Equally
useful, as illustrated in Fig.~1,
the Orion ridge presents a face-on geometry viewed from earth with its UV-illuminated
surface on the near, earth-facing side.  Thus, every line of sight probes a column
of gas from its UV-illuminated surface to \av's in excess of 30 magnitudes in some cases.
Those species whose abundance peaks near the cloud surface would be expected to 
exhibit relatively little variation in the integrated intensity of their optically
thin emission between lines of sight whose depth extends beyond the surface layers.
Conversely, any optically thin emission from species whose abundance 
rises to a near-constant value throughout the cloud would be 
expected to scale with the line-of-sight column density.
Thus, by measuring the correlation between the observed \water\ integrated intensities 
and the optically thin
integrated intensities of both near-surface and volume-tracing species, it is possible
to constrain the depth dependence of the water-vapor emission.

In \S$\,$2 we review the \swas\ and Five College Radio Astronomy Observatory 
(FCRAO) observations used in this study and, in \S$\,$3, we present the results.
In \S$\,$4 we describe the role of line optical depth effects and depletion along with 
two approaches used to analyze the data.  In \S$\,$5 we
discuss the results and implications for our understanding of the water
distribution in dense molecular clouds.

\vspace{2mm}

\section{\bf Observations}

\vspace{-1.1mm}

The observations reported here were obtained with \swas\ and the 
FCRAO 14-m telescope.  \swas\ utilized a 68 $\times$ 54-cm off-axis
primary mirror coupled to two heterodyne receivers and a 1.4~GHz
bandwidth acousto-optical spectrometer (AOS) backend.  \swas\ was able to
observe either the \oxy, CI, \ico, and \water, or the
\oxy, CI, and \iwater\ lines simultaneously, and the AOS
provided the equivalent of 210\kms\ of baseline per spectral line.  
To ensure that the \swas\ lines were centered in the
AOS, regardless of the source \vlsr\ within the Galaxy, the receivers
possessed a tuning range of $\pm\,$182\kms\ at 490~GHz
and $\pm\,$164\kms\ at 553~GHz, in commandable steps of 7.3 and
6.6\kms, respectively.  The full-width-at-half-maximum (FWHM) for
the \swas\ beams were measured on orbit to be 3$.\!^{\prime}$5
$\times$ 5$.\!^{\prime}$0 at 490~GHz and 3$.\!^{\prime}$3
$\times$ 4$.\!^{\prime}$5 at 553~GHz, in good agreement with
the predictions of diffraction theory for the \swas\ telescope
with an 11$\:$dB Gaussian edge taper.  Strip scans of Jupiter
$\pm\,$10\amin\ across the beam minor axes and $\pm\,$12\amin\
across the beam major axes confirmed: (1) that the beams were symmetrical,
with no evidence for vignetting or other distortions; (2) that the
beam centers were spatially co-aligned to within 5\asec, or 
about 1/40 th of the FWHM of the minor axis of the 553~GHz beam; 
and, (3) the results of pre-launch instrument testing which showed
that the highest sidelobe was suppressed by $\sim\,$-17$\:$dB 
with all other sidelobes below -30$\:$dB out to 15\amin\ from the
beam centers (the limit of these measurements).

The \swas\ maps cover a grid of regular spacing of 3.2 arcminutes, 
corresponding approximately to the angular diameter of the minor axis
of the \swas\ beam at 557~GHz, the frequency of the othro-\water\ ground-state transition.
All of the \swas\ observations reported here were conducted by
nodding the entire observatory.  Because there was no change in
the optical path between on-source and off-source reference
observations, the spectral baselines were generally very flat, requiring no more than
a first-order fit to the baseline to produce good-quality continuum-subtracted spectra. 
Spacecraft nodding also ensured that good reference positions were
always used; spatial positions up to 3~degrees in any direction
from the on-source position for each source were selectable and
were chosen to coincide with the closest position exhibiting no
detectable $^{12}$CO $J = $1$\,-\,$0 emission. 
On-orbit tests indicated that the receiver-AOS
system was radiometrically very stable; measurements
demonstrated that on-source integration times of $\sim\,$200~hours
continued to exhibit radiometric performance -- i.e., spectral noise
$\propto\;$1$/\sqrt{\rm time}$ -- in both receivers.  In addition,
\swas\ \water\ spectra of Orion BN/KL obtained 182 days apart were
reproducible within the noise.
The \swas\ \water\ map of Orion was obtained during several periods
of source availability between 20 December 1998 and 8 October 2003.

Between February and May 1999, the FCRAO 16-element SEQUOIA 
array receiver was used to obtain maps of the $^{12}$CO and \ico\ emission
(cf.~Plume~\etal\ 2000).
Between January and June 2004, the FCRAO 32-pixel SEQUOIA array receiver 
(Erickson~\etal\ 1999) was used
to obtain maps of the emission from C$_2$H, HCN, N$_2$H$^+$, CH$_3$OH,
C$^{18}$O and CN.  In April 2005, further observations were obtained in
C$_2$H and N$_2$H$^+$ repeating regions in Orion where the
emission was weak.  The spectral lines observed by the FCRAO are
summarized in Table~1.  The region mapped in Orion
covered the full spatial extent of the \swas\ H$_2$O observations.  
For all observations, the data were obtained using an On-The-Fly observing technique.  
SEQUOIA has the capability of observing two frequencies simultaneously 
which we utilized in mapping the six molecular species.  The spectrometer
for each pixel was a digital autocorrelator with a bandwidth of 50~MHz 
and 1024 spectral channels per pixel leading to a channel spacing that
varied from 0.17 to 0.13 km s$^{-1}$, depending on the line frequency.
The observations were first resampled to form maps 
with data spaced by 20\asec.   The FWHM beam 
size of the FCRAO telescope varies from approximately 46\asec\ at
the CN line frequency to approximately 60\asec\ at the C$_2$H line 
frequency.  For the comparison with \swas\ data, the FCRAO
observations were further convolved with a Gaussian function to 
form spectra with a FWHM angular resolution
of 3.9\amin, the geometric mean of two axes of the elongated \swas\ beam 
at the \water\ line frequency, each centered on the locations of the \swas\ observations 
and spaced by 3.2\amin.   The main beam efficiency, $\eta_{\rm mb}$, of the FCRAO antenna
varies from approximately 0.45 (at 115~GHz) to 0.50 (at $\sim\,$ 100~GHz).
The main beam efficiency for \swas\ was 0.9.

The line profiles vary for different tracers and different sources.
For example, in the center of Orion, there is a prominent outflow component
that contributes significantly to the total intensity. There are also
tracers with multiple hyperfine components, such as CN, C$_2$H, HCN, N$_2$H$^+$.
To better recover the intensity of the relatively quiescent gas of interest
here, one or more Gaussian components have been fitted to each spectrum. 
When multiple hyperfine components were present, the fitting was restricted
by fixing: (1) the relative spacing between peaks to correspond to the
known spectral separation between hyperfine components; and (2) the common
linewidths of each component. For example, to fit the HCN emission, we allowed
the LSR velocity of the main component to be a free parameter, $v_0$, and
required that the other two components be centered at
$v_0-$7.064\kms\ and $v_0+$4.842\kms, respectively, and that all components
have the same linewidth.  In this way, the hyperfine components were 
treated as correlated Gaussian peaks to best recover the total 
line flux.

\vspace{2mm}

\section{\bf Results}

\vspace{-1.1mm}

The results of our mapping efforts are shown in Figs.~2 -- 6.  
Fig.~2
shows a portion of the central ridge of the Orion Molecular Cloud traced by the 110.2~GHz \ico\ 
$J =\,$1--0 transition.  The irregularly-shaped area outlined in grey encompasses the
region mapped in the 556.9~GHz 1$_{10}$--1$_{01}$ ground-state ortho-\water\ transition
by {\em SWAS}.  The smaller grey square centered on ($\Delta\alpha$, $\Delta\delta)\:=\:$(0, 0)
shows the area affected by the
strong outflows from BN/KL and IRc2.  Because the gas associated with the outflow shocks
possesses temperatures, densities, and chemical abundances distinct from the surrounding quiescent 
material (cf.~Blake~\etal\ 1987), data from within this area 
are excluded from the following analysis.  
Fig.~3 shows the central
region of the ridge mapped with 46\asec\dash 60\asec\ spatial resolution in six of the seven species
observed using FCRAO (the 115.3~GHz $^{12}$CO $J =\,$1--0 map is not shown since its 
emission is optically thick within most of the area mapped by \swas).  As in Fig.~2, 
both the region mapped 
in \water\ by {\em SWAS} and the excluded shock-affected area are outlined.

Fig.~4 shows the \swas\ 556.9~GHz 1$_{10}$-1$_{01}$ ground-state  
\water\ integrated intensity map of 
the ridge along with the spectra upon which the map is based.
Figs.~5 and 6 show the 492.2~GHz CI $^3$P$_1 - ^3$P$_0$
and 550.9~GHz \ico\ $J =\,$5--4 integrated intensity maps of the ridge, respectively, also 
obtained using {\em SWAS}, as well as an expanded view of the area mapped deeply in \water.

A total of 86 spatial positions along the ridge were observed by {\em SWAS} with sufficient
total integration times to either obtain convincing detections of \water\ emission or set meaningful
upper limits to this emission.  A considerably larger number of ridge positions were observed using 
{\em SWAS} for which the total integration times were less, but nonetheless sufficient to obtain
good signal-to-noise spectra of the stronger CI and \ico\ $J =\,$5--4 emission
(Plume~\etal\ 2000).  These
shorter-integration-time measurements were used to construct the extended (beyond the water-map)
CI and \ico\ maps shown in Figs.~5 and 6.  Thus,
with the exclusion of the nine shock-affected lines of sight surrounding position (0,0), there
remain 77 spatial positions for which integrated intensities were obtained for all species
and which form the basis of the analysis of the quiescent gas.  As noted in \S2, the spatial grid
of beam positions and the beam size -- synthesized in the case of the higher spatial resolution
FCRAO maps -- are the same for all species.

\vspace{0.1mm}

\section{\bf Analysis}

\vspace{0.2mm}

\subsection{\em Line Optical Depth Effects}

\vspace{-0.8mm}

{\em H$_2$O}:~~Optical depth effects can lead to an underestimate of the total water 
column density along
a given line of sight, creating the appearance that water is a surface tracer when it
is not.  Such can be the case for water under certain restricted conditions
(cf.~Poelman, Spaans, \& Tielens 2007).  Is this the case here?
The 1$_{10}-$1$_{01}$ transition of \nwater\ has a high critical density,
$\sim\,$8$\times$10$^7$~\cmc\ at 30~K, and is expected to have a high optical
depth for even a relatively low ortho-\water\ column density.  Thus, line
trapping plays an important role in the excitation of this transition.  For
large optical depths, the ``effective critical density" is $A_{u\ell}/(C_{u\ell} \tau_o$),
where $A_{u\ell}$ is the spontaneous emission rate, $C_{u\ell}$ is the collisional 
de-excitation rate, and $\tau_o$ is the line-center optical depth.  For densities
less than the effective critical density, line photons may scatter multiple times
but will eventually escape the cloud.  In this limit, the line radiation is 
``effectively optically thin."

We examine the question of whether the observed \water\ lines toward
the Orion ridge are effectively thin in two ways.  First, we compute the emergent
\water\ 1$_{10}-$1$_{01}$ line flux for a set of densities and water
abundances representative of the Orion ridge.  These results are shown 
in Fig.~7.  The line fluxes were computed under the large velocity
gradient (LVG) approximation (see Neufeld \& Melnick 1987, 1991);
the collisional rate coefficients with o-\mh\ and p-\mh\
for the lowest 45 \water\ rotational energy levels, corresponding to a maximum upper-level
temperature of $\sim\,$2000$\:$K,
are those reported by Faure~\etal\ (2007), the first five levels
of which at 20$\:$K are those reported by Dubernet~\etal\ (2006).
In order to bound the range of
likely \mh\ ortho-to-para ratios (OPR), which is presently unknown, results are
given for the LTE OPR value at 30$\:$K as well as an OPR value of 3.
Among the
77 spatial positions considered here, the average measured \water\ line width is 
3.9\kms; we assume a line width of 3.5\kms\ in our
calculations.  Finally, we adopt a slab geometry which, because it yields the lowest
escape probability for a given line-center optical depth, is the most conservative assumption;
a polynomial fit to the exact expression for the photon escape
probability from a plane-parallel emitting region (Hummer \& Rybicki 1982) is used.

The maximum \water\ integrated antenna temperature, $\int T_A dv$, among the 77 positions
considered is 3.26 K km~s$^{-1}$.  As can be seen in Fig.~7, 
for integrated intensities below the maximum observed, the line flux increases 
with column density very nearly linearly, as expected for optically thin emission;
the deviation from linear behavior is less than 20 percent in all cases.  Finally, similar analyses
for gas temperatures of 20$\:$K and 50$\:$K (not shown) support the conclusion that
the \water\ 1$_{10}-$1$_{01}$ emission is effectively thin.

This result is consistent with that of Linke~\etal\ (1977), who
showed that if the main-beam antenna temperature, $T_{\rm mb}$, 
satisfies $T_{\rm mb}\,\ll\,(h\nu/4k)\,{\rm exp}(-h\nu/kT)$, then collisional
excitation of the upper 1$_{10}$ level always results in a photon that 
escapes the cloud.  At a kinetic temperature, $T$, of 30$\,$K, the \water\ 
556.9~GHz line is effectively thin if $T_{\rm mb}<\:$2.7~K, or the antenna temperature
is less than 2.5$\,$K (for $\eta_{\rm mb} =\:$0.9).  
The peak observed antenna temperature for the \water\ emission from the
Orion ridge is $\sim\,$0.4~K, after correction for the \swas\ main-beam efficiency.
Thus, unless the beam filling factor for the \water\ emission is much less
than 0.16, which appears unlikely given the distribution of gas shown in 
Figs.~2--6, the \water\ emission is effectively optically thin.
Future \water\ 556.9~GHz mapping observations toward the Orion ridge
using the {\em Herschel Space Observatory} should be able to further test this assumption.

{\em $^{13}$CO}:~~We are also interested in knowing whether the observed \ico\ emission provides a
good measure of total cloud depth.  To assess this, we compute the \ico\ column
density required to achieve a line optical depth of 1 in the $J\:=\:$1$\,$--$\,$0
transition using the RADEX LVG code (Van der Tak~\etal\ 2007),
the Einstein A-coefficients and collision cross-sections from the Leiden
Atomic and Molecular Database (Sch\"{o}ier~\etal\ 2005) for the lowest 40
rotational energy levels,
and the assumption of no CO freeze-out.  Line
widths of 1.5 and 3\kms\ are assumed;  the average observed \ico\ $J\:=\:$1$\,$--$\,$0
line FWHM is 2.8\kms.  As shown in Fig.~8, for a gas temperature of 30~K, 
the \ico\ $J\:=\:$1$\,$--$\,$0
transition should be optically thin in most cases of interest here.  
However, because a few lines of sight have a \ico\ line width less than 1.5\kms,
or may have a temperature less than 30~K, we choose to reference our measurements
against the \cio\ $J\:=\:$1$\,$--$\,$0 transition 
to ensure a measure of the total cloud column density without concern for optical depth effects.

{\em HCN, CN, C$_2$H, and N$_2$H$^+$}:~~Several of the molecules observed
have distinct hyperfine structure
that can be used to estimate the optical depth of these lines.
In the $J\!=\,$1\dash 0 transition of HCN, we observe the F=1-1, F=2-1 and F=0-1
hyperfine lines; in the $J\!=\,$1\dash 0 transition of N$_2$H$^+$, we observe the 
F$_1$=1-1, F$_1$=2-1 and F$_1$=0-0 hyperfine lines; 
in the $N\!=\,$1\dash 0, J=1/2-1/2 transition of
C$_2$H, we observe the F=1-1 and F=0-1 hyperfine lines; and, in the $N\!=\,$1\dash 0,
J=3/2-1/2 transition of CN, we observe the F=3/2-3/2, F=1/2-1/2,
F=5/2-3/2 and F=3/2-1/2 hyperfine lines.  If we assume that the
hyperfine lines are populated according to LTE, we can use the
observed hyperfine ratios to determine the optical depth in the
strongest hyperfine component.  The greatest leverage on the optical
depth comes from the ratio of the strongest to the
weakest hyperfine line; for spectra with good
signal-to-noise ratios, relatively accurate optical
depths can then be derived.  For those HCN, CN, C$_2$H, and N$_2$H$^+$ spectra 
with good measures of all hyperfine components,
we find that more than 90\% of these spectra are consistent with line-center 
optical depths of less than 1
and, in no instance, was a line-center optical depth greater than 1.5.

% $<$1 in 64 of 65 positions in HCN (98.5\%)
% $<$1 in 49 of 53 positions in CN (92.5\%)
% $<$1 in 38 of 39 positions in C2H (97.4\%)
% $leq$1 in 17 of 19 positions in N2H+ (89.5\%)

\vspace{0.5mm}

\subsection{\em CO Depletion}

\vspace{0.1mm}

The \cio\ integrated intensity can be used as a reliable measure of cloud depth
only if CO remains undepleted throughout the column of
gas observed.  The gas-phase CO abundance is depleted mainly in two ways:

First, it is assumed that all CO that strike dust grains when the grain temperatures are 
$T_{\rm gr}\:\simlt\:$20$\:$K will stick to the surface and be removed from the gas phase.
The timescale for this process is 
$\sim\:$6$\,\times\,$10$^{4}\,$[5$\times$10$^3$~\cmc/\nh] (30$\:$K$\,/\,T$)$^{1/2}$
years (cf.~Hollenbach~\etal\ 2009), thus leading to 
the rapid freeze-out of CO unless subsequently desorbed.

Second, in regions where grain temperatures are above the sublimation temperature of CO-ice,
but below that of \water-ice, i.e., 20$\:$K$\,<\,T_{\rm gr}\,<\,$90$\:$K,
CO can still be depleted through reactions with He$^+$ and
the continuous removal of elemental O from the gas-phase.
Specifically, He$^+$ created by cosmic rays can react with CO to produce C$^+$, O, and He.
The O thus produced can react to reform CO in the gas-phase or form \water\ in the gas-phase
or on grain surfaces.  Whether formed in the gas-phase or on grains,
most of the \water\ will eventually end up on 
grain surfaces, where it will remain unless desorbed.  In the absence of significant FUV 
photodesorption or cosmic-ray desorption, this process results in a steady decrease
in the gas-phase elemental oxygen abundance and the significant depletion of CO in 
about 10$^6$ years.

It is well established that CO suffers strong depletion in the central
parts of dense low-mass cloud cores, such as B68 (Bergin~\etal\ 2006),
L1544 (Caselli~\etal\ 1999), L1498 (Willacy, Langer, \& Velusamy 1998)
IC 5146 (Kramer~\etal\ 1999), and the Taurus molecular cloud (Pineda~\etal\ 2010).  
Given the low dust temperatures measured toward the cores of these regions, it is 
assumed that direct CO freeze-out onto grains is primarily responsible for the 
observed depletion.  

Is CO depletion significant within the Orion ridge?  There are three reasons to believe that it
is not a factor here.  First, infrared and submillimeter observations toward the Orion ridge are
best fit by dust temperatures between about 20$\:$K and 30$\:$K (e.g., Johnstone \& Bally 1999;
Mookerjea~\etal\ 2000), making it unlikely
that direct freeze-out of CO onto dust grains is occurring.  Second, unlike in cold cores,
Zinchenko, Caselli, \& Pirogov (2009) find no evidence for CO depletion 
within other regions of high-mass 
star formation studied -- i.e., W3, DR21, S140, S187, and S255.
Third, the ratio of N$_2$H$^+$ to \cio\ integrated intensities toward the Orion ridge
varies only by about a factor of two over the full range of observed \cio\ integrated 
intensities.  Since negligible N$_2$H$^+$ depletion is observed toward colder regions of
comparable density (see Tafalla~\etal\ 2004, and references therein),
particularly for depths into the cloud corresponding to the visual extinction range 
of greatest interest here, i.e., \av$\,<\,$20 (e.g., Bergin~\etal\ 2002), the
N$_2$H$^+$ to \cio\ ratio would be expected to increase by several orders of magnitude 
if CO depletion were significant.  The absence of any substantial
CO depletion may suggest that the age of the Orion ridge
is less than the timescale for CO depletion by He$^+$ destruction.  Thus, throughout the
remainder of this paper, we assume that CO is undepleted within the portion of the ridge observed 
in \water\ by \swas.

\vspace{0.5mm}

\subsection{\em Abundance Profiles}

\vspace{0.1mm}

Because of the face-on appearance of the Orion ridge and its relatively strong 
extended emission, which permits many independent spatial samples,
this source offers a particularly good opportunity to study observationally the distribution 
of water vapor in dense molecular clouds.  To do so, we seek to examine
correlations between gas-phase \water\ and a number of other species whose distribution
with depth is believed to be well understood.  With a face-on appearance,
one method for investigating the line-of-sight distribution of a species is
to plot its integrated intensity versus that of \cio\ (for the same
spatial positions), where the optically thin \cio\ $J =\,$1$\,-\,$0 emission serves 
as a proxy for the total column thickness of a given line of sight. 
Relating the \cio\ $J =\,$1$\,-\,$0 integrated intensity to the total \cio\
column density, $N$(C$^{18}$O), can be approached in two ways.

First, by assuming the \cio\ emission is optically thin and the background radiation 
terms can be ignored, it is possible to derive a simple analytical relation between
the column density in the upper $J=$1 state of the transition and
the integrated intensity of the line (in K km s$^{-1}$), corrected
by main-beam efficiency ($\sim\,$0.5 for FCRAO at the \cio\
line frequency):

\vspace{-1.4mm}

\begin{equation}
N(J=1)~=~3.8 \times 10^{14}~\times~\int\!{T_R({\rm C^{18}O})\,dv}~~~~{\rm cm^{-2}},
\end{equation}

\vspace{1.0mm}

\noindent where $T_R$ is the radiation temperature 
($= T_{\! A}^{\:*}/$main-beam efficiency).

\noindent Using a standard partition function, this gives the following
expression for the total \cio\ column density:

\vspace{-4.3mm}

\begin{equation}
N({\rm C^{18}O})~=~4.8 \times 10^{13}~T/\left[e^{(-5.27/T)}\right]~\times~
\int\!{T_R({\rm C^{18}O})\,dv}~~~~{\rm cm^{-2}},
\end{equation}

\vspace{3.0mm}

\noindent where $T$ is the gas temperature.

Second, using the LVG approximation, assuming an \mh\ density of 10$^5$~\cmc, and 
including all of the background radiation terms, we compute the \cio\ 
integrated intensity for a range of column densities and temperatures. 
Fitting these data, we obtain the relation:

\vspace{0.3mm}

\begin{equation}
N({\rm C^{18}O})~=~1.9 \times 10^{14}\,~T^{\,0.65}~\times~\int\!{T_R({\rm C^{18}O})\,dv}~~~~{\rm cm^{-2}}.
\end{equation}

\vspace{3.3mm}

\noindent This expression is accurate for temperatures between
15 and 100~K and column densities where the emission is optically thin.
As shown in Fig.~9, the two expressions are in good agreement,
particularly over the range of temperatures most applicable to the Orion ridge --
i.e., 20$\,$--$\,$40~K (cf.~Ungerechts~\etal\ 1997).
At $T\,=\,$30~K, the optically thin criterion is satisfied if the
maximum column density divided by line width (in km s$^{-1}$) is 
$N$(C$^{18}$O)/$\Delta v <\;$3$\,\times\,$10$^{16}$~s$\:$km$^{-1}\:$cm$^{-2}$.  
Since the measured
\cio\ $J =\,$1$\,$--$\,$0 integrated intensities are all less than 4 K km s$^{-1}$,
implying $N$(C$^{18}$O)$\;\simlt\;$7\ti 10$^{15}$ cm$^{-2}$, and the
\cio\ line widths are all greater than 0.95 km s$^{-1}$, the optically thin assumption
is justified for the lines of sight considered here.

The relation between the \cio\ and H$_2$ column densities is
best established in dark clouds and is based on extinction
determinations from the 2MASS data.  For example,
Kainulainen, Lehtinen, \& Harju (2006)
examine the ratio of $N$(C$^{18}$O) and \av\ in Chamaeleon I 
and III-B using 2MASS and SEST data.
Expressed in terms of the total visual extinction, \av,
averaging the two clouds presented in their paper
yields approximately:

\vspace{-3.5mm}

\begin{equation}
A_V~=~5 \times 10^{-15} N({\rm C^{18}O})~+~2.3~=~
0.95~T^{\,0.65}~\times~\int\!{T_R}({\rm C^{18}O})\,dv~+~2.3~~~~{\rm mag.}
\end{equation}

\vspace{3.3mm}

\noindent The offset is due to extinction of the surface layers where the gas-phase 
carbon is C$^+$ or CI, and not CO, and is uncertain and appears to vary
from cloud to cloud (see summary by Harjunp\"{a}\"{a}, Lehtinen, \& Haikala 2004).
Ignoring the offset, this relation leads to an abundance ratio of 
$N$(C$^{18}$O)$\,$/$N$(H$_2$)$\,\sim\;$2$ \times $10$^{-7}$.  
The depth into the cloud, \av, measured in visual magnitudes of extinction, 
as a function of \cio\ integrated intensity for a range of assumed
temperatures is shown in Fig.~10.  These results
are in good agreement with the previous study of
Lada~\etal\ (1994).

Figs.~11 and 12 show plots of the ratio of the
H$_2$O, C$_2$H, HCN, CN, CI, \ico\ $J\,=\,$1$\,-\,$0, and N$_2$H$^+$
integrated intensities to those of \cio\ $J\,=\,$1$\,-\,$0 as a function of
the \cio\ $J\,=\,$1$\,-\,$0 integrated intensity.  The corresponding depth into 
the cloud, in visual magnitudes, is shown along the top axis of each plot.
For these values, Eqn.~(4) is used assuming $T =\:$30$\:$K.
For the near-surface depths of particular interest here, corresponding to \cio\ 
integrated intensities less than about 1.5~K km s$^{-1}$, 
the \av\ derived from the \cio\ intensity is relatively insensitive to the assumed 
temperature.

Because the emission from most species toward BN/KL is strongly affected by the outflow, the
data corresponding to the ($\Delta\alpha, \Delta\delta)\:=\:$(0, 0) and 
surrounding 8 positions are not included in these plots.  
To better reveal any trends (by reducing the scatter in the 77 data points), 
the ratio values have been co-averaged in bins of \cio\  $J=\,$1-0 integrated intensity
of width 0.2~K km s$^{-1}$ in the $x$-axis.
The plotted $y$-value within each bin is the weighted mean,
$\mu\,=\,\Sigma(y_i/\sigma_i^2)/\Sigma(1/\sigma_i^2)$,
of the $i$ data points lying within that bin, 
where $y_i$ is the ratio of the integrated intensity, $I$, of species $a$ to species $b$ for point $i$,
i.e., $y_i=I_{a, i}/I_{b, i}$, and 
$\sigma_i\,=\,\left[y_i^{\;2}\,\left(\sigma_{a,i}^{\;2}/I_{a, i}^2\,+\,\sigma_{b,i}^{\;2}/I_{b, i}^2\right)\right]^{1/2}$,
where $\sigma_{a,i}$ and $\sigma_{b,i}$ are the 1$\sigma$ uncertainties associated with
the $i$-th integrated intensity for species $a$ and $b$, respectively.
The 1$\sigma$ $y$-value error bars represent the uncertainty of the mean,
$\sigma_{\mu}\,=\,\left[1/\Sigma(1/\sigma_i^2)\right]^{1/2}$.  Though barely visible in these plots,
the 1$\sigma$ error bars representing the $x$-value dispersion within each bin 
are also shown.

The results fall broadly into two categories -- i.e., those species that exhibit an increase in their
integrated intensities relative to \cio\ toward lower \av's and one that shows an increase
in this ratio with depth.  Specifically, C$_2$H, CN, HCN, and 
CI all show a steady rise in the observed intensity ratio toward the cloud surface,
with the possible indication that the CN and HCN profiles
subsequently decrease at \av$\:<\:$5.
Conversely, the ratio of N$_2$H$^+$ to \cio\ integrated intensities appears to  
increase with depth.

One measure of whether these plots convey an accurate picture of the
abundance profiles is provided by the observed profile of \ico/\cio\ integrated intensities,
shown in Fig.~12.  Assuming the observed \ico\ $J\:=\:$1$\,-\,$0 line is
optically thin and depletion of CO is not significant along the ridge,
the \ico/\cio\ intensity ratio deep in the cloud is expected to reflect the
isotopic ratio of $^{16}$O/$^{18}$O of 500 and $^{12}$C/$^{13}$C of between 43
(Hawkins \& Jura 1987; Stacey~\etal\ 1993; Savage~\etal\ 2002)
and 65 (Langer \& Penzias 1990), i.e., \ico/\cio$\;\simeq\;$8-12.  The observed 
\ico/\cio\ intensity ratio deep in the cloud is in good agreement with these values 
and, thus, provides reason to  believe the inferred profiles are descriptive of 
the actual profiles.

Fig.~11 clearly shows an increase in the \water/\cio\ intensity ratio 
at \av$\:<\:$15, with a steady rise toward the cloud surface.  Because the increase is evident
between \av\ $\sim\,$5 and 15, where the \cio\ abundance is predicted to be approximately
constant, the inferred increase in the \water\ emission toward
the cloud surface appears to be real.

\vspace{0.5mm}

\subsection{\em Principal Component Analysis}

\vspace{-1.1mm}

A second method for studying the correlations between species involves use
of multivariate analysis referred to as Principal Component Analysis (PCA).  
PCA's goal is to find, among linear combinations of the data variables, a
sequence of orthogonal, or completely uncorrelated, factors that most efficiently
explain the differences in the data.
Details of this method are provided elsewhere
(cf.~Ungerechts~\etal\ 1997, for applications to 
astronomical mapping data) and
will not be repeated here, except to note that the PCA approach provides a
useful and compact means for quantifying the commonality between maps made
in different transitions.

In PCA, we attempt to explain the total variability of $p$ correlated variables through 
the use of $p$ orthogonal principal components (PC). The components themselves are 
merely weighted linear combinations of the original variables such that PC$\:$1
accounts for the maximum variance in the data of any possible linear combination,
PC$\:$2 accounts for the maximum amount of variance not explained by PC$\:$1
and that it is  orthogonal to PC$\:$1, and so on.  Even though use of all $p$ PC's permits the full
reconstruction of the original data, in many cases the first few PC's are sufficient to
capture most of the variance in the data.  Thus, we can express each observed map 
(to within the noise) as a different linear combination of just two or three maps 
(i.e. two or three principal components).

To ensure that the analysis gives equal weighting to each line -- versus allowing the
brightest lines to dominate the analysis -- the integrated intensities for each spectral
line have been mean subtracted and divided by the standard deviation.  
In addition, to reduce the variance
due to excitation and varying line emissivity, only those transitions 
with a critical density greater than 10$^4$~\cmc\ (see Table~1) are included in the PCA.  
The results are shown in Fig.~13.

The interpretation of these results is straightforward.  The top panel in Fig.~13 plots
the fraction of the total variance in the data captured by each principal component.
For the data considered here, 90\% of the total variance between species is accounted for with 
principal components one and two, and 96\% of the total variance is accounted for 
with the addition of principal component three.  

The lower two panels in Fig.~13 plot the coefficients for the first and second, and
second and third, principal components, respectively.  Because almost all of the variation
in the data is in principal components one and two, the bottom left panel
is most relevant.  There are two key elements in the plot to note: (1) the degree to which
each vector approaches the unit circle; and, (2) the clustering of vectors.
Because the principal components are normalized such that the quadrature sum of the 
coefficients for each species is unity, the proximity of the points to the circle of unit
radius is a measure of the degree to which any two principal components account for
the total variance in this sample.  Consequently, the closeness of all the points to the
unit circle in the lower left panel is a reflection of the fact that these two principal 
components contain almost all of the variance in the data, as noted above.

The degree of clustering of the vectors is a measure of their correlation.
As noted by Neufeld~\etal\ (2007),
in the limit where two points actually lie on the unit circle, the cosine of the angle
between these points is their linear correlation.  Thus, points that coincide
on the unit circle (i.e., $\Delta\theta\,=\,$0\ddeg) would indicate perfectly correlated data,
whereas points on opposite sides of the circle (i.e., $\Delta\theta\,=\,$180\ddeg) would indicate 
perfectly anticorrelated data.  Thus, the plot of PC1 versus PC2 quantifies what appears
evident in Figs.~11 and 12, namely that the ground-state 
ortho-\water\ emission is well
correlated with the CN, HCN, and C$_2$H emission.  The \water\ emission is also
well correlated with the \ico\ $J = $5--4 emission.  Since the upper level of this
\ico\ transition is 79$\:$K above the ground state, it too is expected to
trace the warmer surface layers of the ridge.  The vectors representing \water\
and N$_2$H$^+$ in this plot show the greatest separation, suggesting no particular
correlation exists between the lines of sight where the \water\ and N$_2$H$^+$ integrated 
intensities are strong.

The bottom right panel in Fig.~13, i.e., PC2 versus PC3, shows the distribution of 
residual variance between the species.  The short vectors (from the 0,0 point) illustrate 
quantitatively the relative unimportance of additional sources of variance between species
beyond those captured in the first two principal components.

\vspace{2mm}

\section{\bf Discussion}

\vspace{-1.1mm}

The distribution of water vapor within a molecular cloud depends upon both the
gas-phase chemistry that forms \water\ and a number of important micro-physics 
processes.  These processes include the 
rate at which oxygen atoms strike dust grains and combine with hydrogen on their
surfaces to form
OH and \water, the rate at which such \water\ 
is removed from grain surfaces by incident UV photons (i.e., photodesorption)
and cosmic rays, and
the rate at which gas-phase \water\ is destroyed by UV photons (i.e., photodissociation).
Because these processes are operative in all quiescent molecular clouds to varying degrees, 
testing models that incorporate the necessary chemistry 
and physics is important.  Water is a particularly good diagnostic
since its gas-phase abundance is sensitive to all of the above processes.  In addition,
water is a molecule of great (and growing) inherent interest.

The need to refine our understanding of water vapor in molecular clouds, and the
motivation for this study, results from two main \swas\ findings.  First, 
it became clear early in the \swas\ mission that more than just gas-phase chemical 
models are needed to explain the inferred \water\ abundances toward quiescent clouds.
In particular, in cold ({\em T}$\,\simlt\,$30~K), dense ($n$(\mh)$\:\simgt\:$10$^4$~\cmc)
clouds, the inferred
abundance of gaseous \water\ is $\sim\,$100 to 1000 times less than the
predictions of ion-neutral gas-phase chemical models and the \oxy\ 
abundance is at least 100 times less than predictions 
(cf.~Neufeld, Lepp, \& Melnick 1995; Snell~\etal\ 2000c;
Goldsmith~\etal\ 2000; Bergin~\etal\ 2000; Melnick 2004).

A post-flight review of the \swas\ data revealed a second important clue.
As shown in Fig.~14, the peak antenna temperatures observed toward
83 giant and dark cloud cores display a relatively small spread in values;
almost 70 percent of the sources observed have a peak antenna temperature 
within a factor of two of 100~mK.
To understand why this is significant, it is useful to consider how the line 
emission scales with the physical conditions.  For effectively thin emission, the
1$_{10}-$1$_{01}$ 556.9~GHz integrated intensity can be expressed as

\vspace{-1.8mm}

\begin{equation}
\int T_A\:dv  =  \eta_{\rm mb}\,{C_{u\ell}}\,\left(\frac{h\nu}{4\pi}\right)\,
\left(\frac{c^3}{2 k \nu^3}\right)\:{\sl x}({\rm o-H_2O})\:N{\rm (H_2)}\,n{\rm (H_2)}\,
e^{-h\nu/kT_K}~~~~{\rm K~km~s^{-1}},
\end{equation}

\vspace{3.7mm}

\noindent where $\nu$ is the line frequency,
%$\eta_{\rm mb}$ is the main beam efficiency
\xwater\ is the ortho-\water\ abundance (relative to \mh),
$N({\rm H_2})$ is the \mh\ column density, and $n{\rm (H_2)}$ is
the \mh\ volume density.  Consequently, in the optically thin limit, 
the integrated intensity increases
linearly with increasing \mh\ column and volume densities, even if
the line center optical depth is large.  Since the measured velocity-resolved 
line widths (FWHM) toward quiescent giant and dark clouds are all within a factor 
of two of 5\kms, the peak antenna temperatures should reflect the spread in  
\xwater, $N({\rm H_2})$, and $n({\rm H_2})$ among the sources.

The relatively small variation in the peak \water\ line intensity between sources 
with more than order-of-magnitude differences in \mh\ column densities (inferred 
from both \ico\ and \cio\ measurements), and \mh\ volume densities 
(inferred from a variety of molecular species) suggests that the ortho-water
column density, i.e., \xwater$\:N({\rm H_2})$, and \mh\
density within the water-emitting region are not particularly
sensitive to the total depth and peak density of the target clouds.  
Such could only be the case if the water-vapor emission originates
predominantly from a zone within the cloud whose density, column density, and \water\
abundance are tightly coupled, with little variation from cloud to cloud.

Ideally, it would be desirable to directly measure the depth dependence of the water
emission.  Regrettably, this is impossible with the face-on appearance of the Orion ridge and attempts
to do so indirectly  -- for example, by modeling the emission from each of the 77 lines of sight 
considered here -- are complicated by often unknown variations in the physical conditions along
the ridge, including the varying incident FUV flux.  Instead, we seek to 
extract the desired information from the ensemble of data, such as whether the 
water-vapor emission correlates better with surface tracers (i.e., species whose depth of 
peak abundance is relatively small and located near the cloud surface), or volume tracers (i.e., 
species whose abundance increases and attains a near-constant maximum with depth into
the cloud).
%In the discussion below, we present the
%predicted depth-dependent abundance profiles from a number of 
%photodissociation region (PDR) models and compare
%these with our Orion observations.  Next, we correlate the observed water vapor emission
%with our other measured species.  Finally, we discuss how these results compare with
%current theory.

Models of PDR's provide the detailed predictions necessary to identify the surface and
volume tracers.  Unfortunately,
the depth-dependent abundance profiles are affected by the
strength of the incident FUV field, which varies along the Orion ridge.  
Stacey~\etal\ (1993)
have modeled the FUV-sensitive 157.74\um\ [C$\,$II] emission from OMC-1 and
find that the value of $G_{\rm o}$, the factor by which the FUV field exceeds the local
interstellar value, 
 at $d$ parsecs projected distance, or 
$\theta_{\rm sep}$ arcminutes angular separation, between the Trapezium cluster 
and a given point along the ridge is given by

\vspace{0.6mm}

\begin{equation}
G_{\rm o} = \frac{8.6 \times\ 10^3}{\left( d^{\;2} + d_c^2 \right)^{1.5}} = 
\frac{8.6 \times\ 10^3}{\left( \left[480\,{\rm tan}(\theta_{\rm sep}/60)\right]^2
+ 0.152 \right)^{1.5}}~~,
\end{equation}

\vspace{4.8mm}

\noindent where $d_c$ is the distance, in parsecs, along the line of sight between the 
foreground Trapezium cluster and the molecular cloud, assumed to be 0.39~pc (see Fig.~1).
Thus, for the ridge positions considered here, \go\ is expected to vary between
$\sim\,$2\ti 10$^4$ and $\sim\,$100 from about 5 to 30 arcminutes from BN/KL, 
respectively.  The presence of a large number of B-stars in the Orion molecular cloud complex, 
in addition to the OB stars in the Trapezium cluster, suggests that the strength 
of the FUV field predicted by Eqn.~(6) far from BN/KL is likely to be a lower limit.

Fig.~15 shows the predicted depth-dependent gas-phase abundance 
profiles for the observed species for 100$\:\leq\:$\go$\:\leq\:$2$\times$10$^3$,
while Fig.~16 shows the predicted profiles for
10$^4 \leq\:$\go$\:\leq\:$2$\times$10$^5$.  It is important to note that these
models assume that \water\ remains in the gas phase throughout the cloud and
does {\em not} freeze out.  As 
discussed earlier, the freeze out of \water\ locks elemental oxygen in ice, reducing 
the gas-phase oxygen abundance and altering the gas-phase chemistry.  If this is the case,
as the current data suggest, then the abundance profiles in Figs.~15
and 16 no longer reflect the actual abundance profiles at depths 
greater than \av$\,\sim\,$6\dash 8 where \water\ begins to freeze out for the 
densities and \go's relevant to the Orion ridge (see Hollenbach~\etal\ 2009).
Nevertheless, the \water\ freeze-out
point lies beyond the depth where all but two of the species observed here -- i.e., CO and 
N$_2$H$^+$ -- peak and, thus, these figures provide some guidance.  (As discussed in
$\S$4.2, it is assumed that CO and N$_2$H$^+$ are undepleted within the region
of interest here.)

Over the broad range of FUV field strengths relevant here, 
species such as C, CN, C$_2$H, and HCN are predicted to reach their peak
abundance within 8 visual magnitudes of the cloud surface and subsequently decrease in
abundance.  Though observations of CI provide
some evidence for emission in molecular cloud interiors 
(e.g., Keene~\etal\ 1985),
its creation via CO photodissociation clearly establishes this species as 
a surface tracer.
C$_2$H and CN have been found to trace the edges of clouds exposed to UV radiation
(Jansen~\etal\ 1995b; Rodr\'{i}guez-Franco~\etal\ 1998).
In general, these simple carbon-based 
radicals can form rapidly via reactions between C and C$^+$ with other simple molecules
(e.g., Sternberg \& Dalgarno 1995).
This will lead CN and C$_2$H to trace regions where 
CI and C$^+$ are abundant (e.g.~the cloud surface).   The formation of HCN is more complex, 
with several contributing pathways, but in general this species can form in abundance in 
regions where CI/CO$\;\sim\:$1, and thus it too can appear in abundance at low to moderate 
extinctions where CI is present.

Alternately, species such as \nco, \ico, \cio, and N$_2$H$^+$ rise in 
abundance between 2 and 8 visual magnitudes of the surface
and maintain near-constant values with increasing depth.  In the case of
N$_2$H$^+$, this molecule requires the {\em a priori} formation of 
N$_2$, which itself forms in a staged process through 
N $+$ OH $\rightarrow$ NO $+$ H and NO $+$ N $\rightarrow$ N$_2$ + O.
These neutral-neutral reactions are not fast enough to compensate for photodissociation, 
leading N$_2$H$^+$ to preferentially appear in abundance only at greater depths.

For CI, CN, HCN, C$_2$H, \ico\, and N$_2$H$^+$, the trends indicated in the
models are reflected in the observed profiles shown in Figs.~11 
and 12.  Because the \cio\ abundance is predicted to
decrease sharply for \av$\:<\:$ 5 (due to FUV photodestruction of the molecule),
it's useful to restrict our consideration to \av$\;\simgt\:$5 where the
\cio\ abundance is predicted to be relatively constant.
Specifically, C$_2$H, 
CN, and CI, all of which have their predicted peak abundance at \av$\:<\:$4 over the full range
of \go, show a steady rise in the observed intensity ratio toward the cloud surface.
Likewise, HCN, which is predicted to peak in its abundance at \av$\;\simeq\;$6$\,-\,$7,
shows an increase in the measured HCN/\cio\ intensity ratio toward these \av's, with perhaps
an indication that the HCN abundance may yet be higher than predicted at \av$\:\sim\:$5. 
Conversely, N$_2$H$^+$, which is predicted to achieve a near-constant abundance at \av$\:\simgt\:$9,
if anything shows a slightly increasing N$_2$H$^+$/\cio\ intensity ratio beyond an \av\ of
$\sim\:$25.  Within a number of other high-mass clouds, the N$_2$H$^+$ abundance 
is observed to increase in regions of lower fractional ionization due to decreased rates of 
dissociative recombination (Zinchenko, Caselli, \& Pirogov 2009). 
This may also explain the rise in the N$_2$H$^+$/\cio\ intensity ratio toward higher \av's 
observed in Orion.

However, the observed \water\ profile is in conspicuous disagreement with the 
predictions of the PDR models summarized in Figs.~15 and 16.
These models, which neglect \water\ freeze-out, predict a
steady increase in the gas-phase water abundance between an \av$\:\sim\:$2 and 10,
reaching a steady-state abundance of $\sim\:$2$\times$10$^{-5}$.  In addition to
predicting a gas-phase \water\ abundance more than two orders of magnitude greater
than observed, the predicted \water\ abundance profile would lead to a tighter
correlation with N$_2$H$^+$, and a reduced correlation with \ico\ $J\:=\:$5$\,-\,$4,
CN, HCN, and C$_2$H, than indicated by the principal component analysis.

It is worth asking whether the correlation between the \water\ emission and the other 
surface tracers reflects little more than the preferential excitation of water vapor in the warmer 
surface layers?  To examine this possibility 
we consider three cases shown in Fig.~17.  
First, we compute the depth-dependent \water\ emission 
resulting from the PDR models of Sternberg \& Dalgarno (1995), 
who considered the chemistry within dense (\nh$\:=\:$5$\times$10$^5$~\cmc) gas 
subject to strong ($G_{\rm o} =\:$2$\times$10$^5$)
external FUV irradiation, but no \water\ freeze-out. 
In their model, \water\ is produced at \av$ =\:$0.6 and 
$T =\:$800$\:$K by the neutral-neutral reactions, 
O$\,+\,$\mh$\:\rightarrow\:$OH$\,+\,$H and 
OH$\,+\,$\mh$\:\rightarrow\:$\water$\,+\,$H.  However, the abundance of \water\ 
in this hot gas layer
is suppressed by the high FUV field which rapidly photodissociates the water.
Deeper into the cloud, i.e.~at \av$\:>\:$5, where the FUV is attenuated and the gas
temperature has dropped to $\sim\,$22$\:$K in their model, a series of gas-phase 
ion-neutral reactions produces a relatively high abundance
of \water\ (\xwater$\,\sim\:$3$\times$10$^{-5}$) which survives
photodestruction.  For the sample case considered, this model would result in more than
90 percent of the water vapor emission arising between \av$\:=\:$5 and 20, and
this fraction would be expected to increase further with increasing cloud depth.  Even for a
line of sight with a total depth equivalent to \av$\:=\:$10, approximately 80 percent
of the water-vapor emission is calculated to arise at \av$\:>\:$5.  Thus, a PDR model
which considers water formed only in the gas phase, and which remains in the gas phase
through the depth of the cloud, does not fit the Orion data.

To assess the case of constant water vapor abundance throughout the cloud volume, we
apply the Sternberg and Dalgarno model above, except the ortho-\water\ abundance
is assumed to be 5$\times$10$^{-7}$ (relative to \mh) and constant with depth.  This value
is at the high end of, but nonetheless consistent with, the range of water abundances 
inferred from \swas\ observations of quiescent clouds, assuming the depths of the
\water\ and \mh\ regions are the same (e.g., Snell~\etal\ 2000c). 
As shown in the middle panel of Fig.~17, most of the water-vapor emission 
originates throughout the volume of the cloud.  
Nonetheless, for clouds of total depth less than an \av\ of about 12, more
than half of the total water emission would originate in the warm surface layers
(i.e., \av$\:\simlt\:$3).  In practice, however, the presence of a 
$G_{\rm o}\:\simgt\:$100 FUV field would destroy 
almost all of the water near the cloud surface, as is evident from the \water\ abundance
profiles in Figs.~15 and 16, and more recently confirmed by
Hollenbach~\etal\ (2009).  Thus, for the case of a constant
water abundance throughout the cloud, the water emission would again be expected to
increase with cloud depth, which it does not.

Third, we consider a temperature and \water\ abundance profile based on the model of 
Hollenbach~\etal\ (2009), except for \nh$\:=\:$5$\times$10$^4$~\cmc\
and $G_{\rm o} =\:$10$^5$.  In this model, the temperature and chemical structure of
a cloud are determined not only by the gas-phase chemistry, but also by the freezing of species
onto grains, simple grain surface chemistry, and desorption (including FUV
photodesorption) of ices.  The resulting gas-phase \water\ abundance is found to peak in 
a region between the cloud surface, where \water\ is photo-destroyed by FUV photons, 
and the deeper interior, where the FUV field is highly attenuated and gas-phase
\water\ depletes rapidly onto grains as frozen water-ice (see Fig.~18).  
Because the sublimation temperature of water-ice 
is high (i.e., $\simgt\:$90$\:$K), it remains on the grains until sputtered off by the
passage of a nondissociative shock, heated by an embedded source, photodesorbed by
FUV radiation, or removed by cosmic rays.  Specifically, this model predicts that 
the gas-phase \water\ and \oxy\ 
lies predominantly between an \av\ of approximately 3 and 8 for \go$\,=\,$1\dash 10$^3$, 
with the peak abundance occurring 
at a depth proportional to ln$\,$($G_{\rm o}/n_H$), where $n_H$ is the gas-phase
hydrogen nucleus number density.  As shown in the right panel of 
Fig.~17, for this scenario (\go$\,=\,$10$^5$) all of the water-vapor 
emission is predicted to 
arise from within a narrow range of depths around \av$\simeq\:$10
corresponding to the peak in the gas-phase \water\ abundance.  For lower values of \go, 
but approximately equivalent densities, the peak gas-phase \water\ abundance shifts 
to lower \av's, as shown in the center and right panels of Fig.~18.
However, unlike the assumptions underlying our analysis of the Orion ridge observations,
the Hollenbach~\etal\ models shown in Fig.~18 depict
steady-state abundance profiles, which include the effects of CO destruction due to He$^+$.
Without CO depletion, these models
underestimate the gas-phase \water\ and CO abundances at high \av.  However, the peak of 
water abundance at intermediate \av\ is likely preserved, consistent
with the observations.

Finally, we note that the observed C$_2$H and N$_2$H$^+$ transitions
have similar excitation energies and critical densities (see Table~1), yet show strikingly 
different depth profiles (e.g., Figs.~11 and 12).  
Thus, the variations in their depth-dependent integrated intensity profiles are not 
the result of excitation conditions.

Limits to the assumption of a simple homogeneous slab geometry, illuminated from 
one side, are also seen in the data.  Specifically the intensity ratios for species such as 
CI, CN, HCN, and C$_2$H
exhibit values at high \av's (e.g., $\simgt\:$15 mag.) that exceed those expected 
on the basis of their low deep-cloud abundances as shown in Figs.~15 and 16.  
This is most likely the result of the Orion ridge being 
somewhat clumpy (cf.~Stacey~\etal\ 1993, and references therein), 
permitting partial penetration of FUV photons deep into the cloud interior.  In addition, 
B stars embedded within the molecular cloud may provide additional FUV flux.  
Thus, emission characteristic of predominately surface tracers can still be generated,
albeit at a lower intensity, well within the cloud interior.  
In addition, if the effects of \water\ freeze-out are added to the models shown in
Figs.~15 and 16,
the deep-cloud abundance of CI, CN, HCN, and C$_2$H may be altered.
This effect notwithstanding, the derived emission profiles for these species follow 
the trends predicted from PDR models, at least up to the depth where \water\ freezes out.

The conclusion that ground-state water-vapor emission is observed to arise primarily 
near molecular cloud 
surfaces is in accord with models, like Hollenbach~\etal\ (2009),
in which the gas-phase water abundance peaks where rates of photodestruction, photodesorption,
and freeze-out balance.  In this model, the distance from
the cloud surface to the peak water-vapor abundance scales as ln$\,$($G_{\rm o}/n_H$) and the
range of depths over which the water vapor is relatively abundant is self regulating and remains
approximately constant.  Consequently, the water-vapor
column density remains approximately constant over a broad range of FUV fluxes and densities,
thus explaining the relatively narrow distribution of observed \water\ 1$_{10}-$1$_{01}$ 
peak antenna temperatures.

This picture is also consistent with the low observed upper limits to the \oxy\
abundance (Goldsmith~\etal\ 2000).  Since the ion-neutral reactions leading to the formation 
of \oxy\ depend upon the abundance of gas-phase O
(via the reaction O $+$ OH $\rightarrow$ \oxy $+$ H), reducing the atomic oxygen abundance
by locking it in water ice suppresses \oxy\ production where water ice becomes abundant -- i.e., 
beyond an \av\ of $\sim\,$2 to 10.  Thus, like water vapor, gas-phase \oxy\ 
is restricted to a relatively narrow zone between where it's photodestroyed near the
cloud surface and where its formation is suppressed by a diminishing supply of atomic oxygen .  
Detailed calculations (e.g., Hollenbach~\etal\ 2009) indicate that 
the resulting \oxy\ column densities are typically between 10$^{15}$ and 10$^{16}\:$\cms, 
consistent with current observed limits.

Finally, evidence supporting an increasing column density of water-ice with depth is provided
by observations of the water-ice band at 3\um\ toward Taurus (Whittet~\etal\ 2001)
and Rho Ophiuchus (Tanaka~\etal\ 1990), 
and the water-ice band at 6\um\ toward Cepheus A East 
(Sonnentrucker~\etal\ 2008).
The onset for water-ice formation toward Taurus, Rho Ophiuchus,
and Cepheus A East is determined to occur at \av's of
3.2, $\sim\,$10, and 2.3 visual magnitudes, respectively,
with the variation due primarily to the different FUV radiation environments for
each cloud.  These observations also confirm the presence of significant amounts 
of water-ice ({\it x}(H$_2$O--ice)$\,>\:$5$\times$10$^{-5}$) 
in the deep interior of dense clouds
(e.g., Nummelin~\etal\ 2001; Boogert~\etal\ 2004; Sonnentrucker~\etal\ 2008).
The detection of CO$_2$ ice with an abundance of $\sim\,$2$\times$10$^{-5}$ 
(Boogert~\etal\ 2004; Sonnentrucker~\etal\ 2008)
further attests to the importance of ices as a repository of oxygen that would
otherwise be available to form gas-phase \water\ and \oxy.

In summary, the results of the observational study presented here show that most of the
water vapor detected toward the Orion Molecular Cloud ridge originates near the cloud
surface, between an \av\ of about 2 and 10.  This finding is in general
agreement with PDR models that consider the effects of photodissociation, grain-surface
chemistry, photodesorption, and freeze-out in addition to gas-phase chemistry.  Future observations
with the {\em Herschel Space Observatory} will allow more detailed follow-up studies of the 
water-vapor distribution in molecular clouds in several ways.  First, {\em Herschel}'s smaller beam
size at 557~GHz -- 40\asec\ versus $\sim\,$230\asec\ for \swas\ -- will permit many more
spatial samples than obtained here, improving the statistics for the type of analysis applied
here.  Second, access to additional ortho- and para-\water\ transitions will enable a more
direct determination of the physical conditions in the water-vapor emitting region.
Combined, these capabilities will allow increasingly stringent tests of our models
of water in molecular clouds.

\clearpage

\bibliographystyle{plainnat}
%\begin{thebibliography}{\itemsep -2mm \parsep -2mm}

\clearpage

\begin{figure}[h]
\centering
\vspace{0.45in}
\includegraphics[scale=0.59]{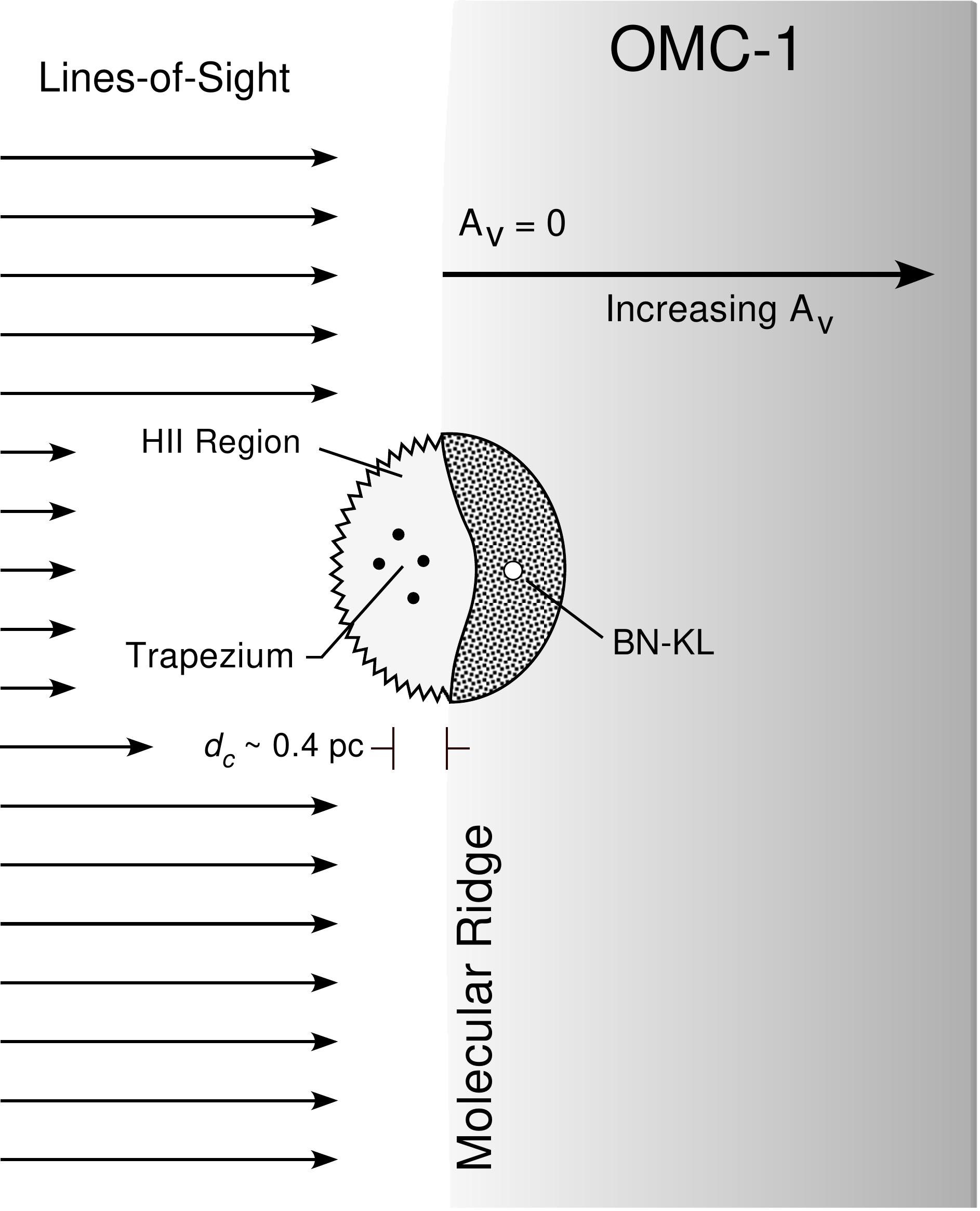}
\renewcommand{\baselinestretch}{0.97}
\vspace{9mm}
\caption{Schematic representation of the lines-of-sight from earth
in relation to the structure of the Orion Molecular Cloud ridge.  The face-on
geometry of the molecular ridge permits a relatively straightforward 
comparison of the distribution of tracers as a function of \av.
The distance between the Trapezium cluster and the ridge, $d_c$,
is noted.}
\renewcommand{\baselinestretch}{1.0}
\label{geometry}
\end{figure}

\clearpage

\begin{figure}[t]
\centering
\vspace{-3mm}
\includegraphics[scale=0.81]{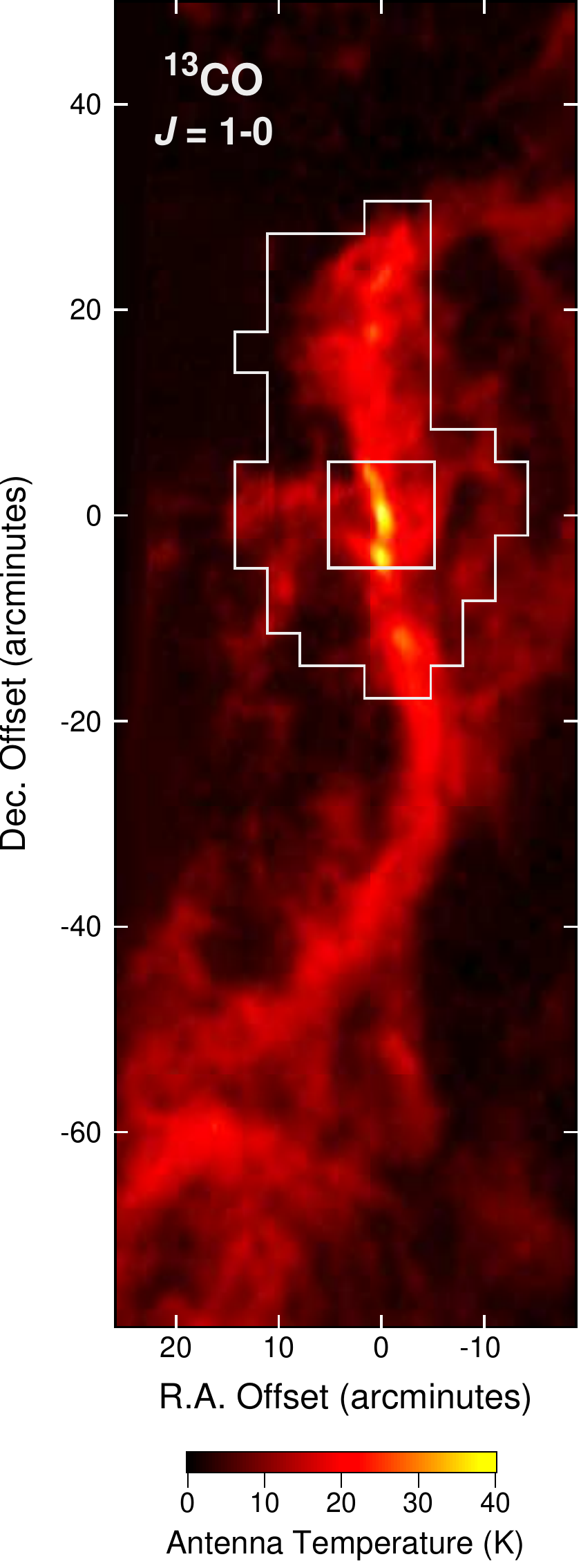}
\vspace{0.2mm}
\renewcommand{\baselinestretch}{0.97}
\caption{Integrated intensity map of the Orion Molecular Ridge made in the \ico\ $J =\:$1-0
110.2~GHz transition using the Five College Radio Astronomy Observatory
(see Table~1).  The beamsize was 47\asec.  The larger grey-outlined region
encompasses the region mapped in the \nwater\ 1$_{10}$-1$_{01}$ 556.9~GHz
transition by {\em SWAS}.  The inner region, denoted by the grey square, includes gas
also subject to strong outflow shocks and is not included in the analysis.
All offsets are relative to $\alpha\,=\,$05$^{\rm h}$35$^{\rm m}$14$^{\rm s}\!\!.$5,
$\delta\:=\,-$05\ddeg 22\amin 37\asec\ (J2000).}
\renewcommand{\baselinestretch}{1.0}
\label{fig:map1}
\end{figure}

\clearpage

\begin{figure}[t]
\centering
\vspace{-3mm}
\includegraphics[angle=90,scale=0.962]{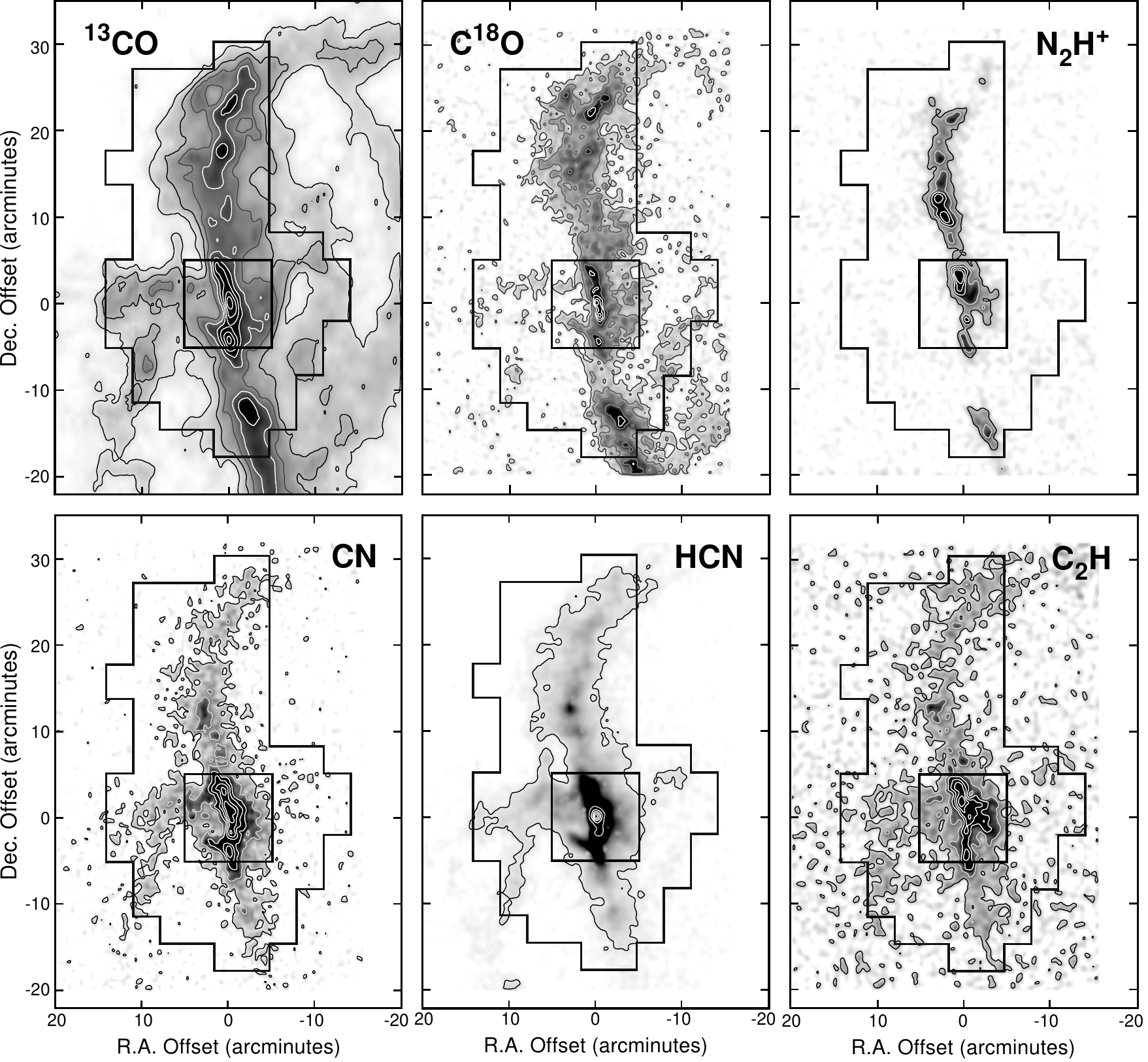}
\vspace{-1mm}
\renewcommand{\baselinestretch}{0.97}
\caption{Maps of the Orion Molecular Ridge obtained with 46\asec\dash 60\asec\ spatial resolution
using the Five College Radio Astronomy Observatory (see Table~1).  The peak integrated 
intensities ($\int\!T_A^{\:*}\:dv$), in K km s$^{-1}$, are:
47.39 (\ico), 6.896 (C$^{18}$O), 14.16 (N$_2$H$^+$), 29.85 (CN), 325.3 (HCN),
6.556 (C$_2$H). Contours superposed on the \ico\ and \cio\ maps are in units of 0.10 
of the peak value, with the peak contour shown being 0.9.  
Contours superposed on the N$_2$H$^+$, CN, HCN, and C$_2$H maps 
are in units of 0.15 of the peak value.  The larger outlined region
encompasses the region mapped in the \nwater\ 1$_{10}$-1$_{01}$ 556.9~GHz
transition by {\em SWAS}.  The inner region, denoted by the square, includes gas
also subject to strong outflow shocks and is not included in the analysis.
All offsets are relative to $\alpha\,=\,$05$^{\rm h}$35$^{\rm m}$14$^{\rm s}\!\!.$5,
$\delta\:=\,-$05\ddeg 22\amin 37\asec\ (J2000). }
\renewcommand{\baselinestretch}{1.0}
\label{fig:map2}
\end{figure}

\clearpage

\begin{figure}[t]
\centering
\vspace{-5mm}
\includegraphics[angle=90,scale=0.76]{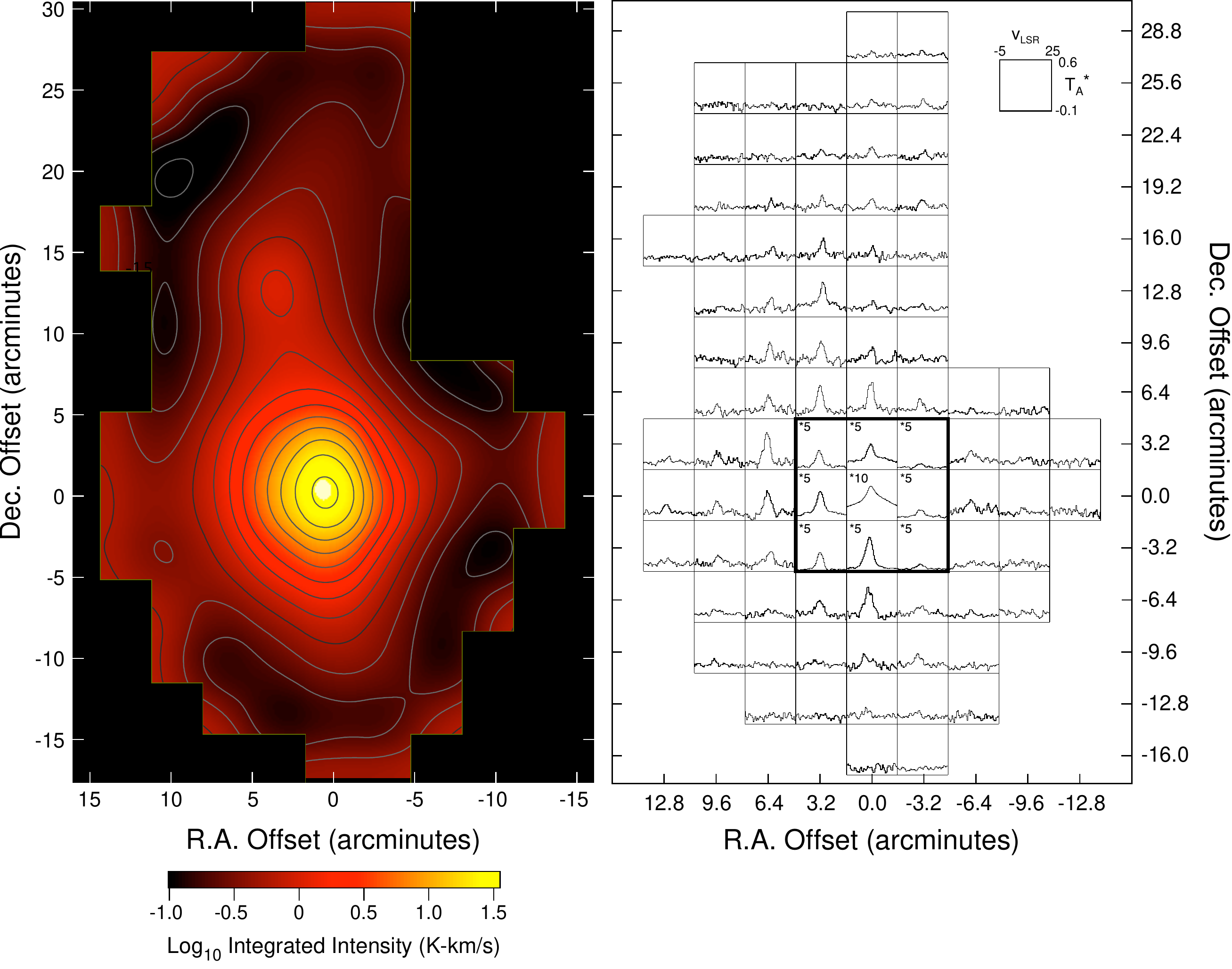}
\vspace{0.1mm}
\renewcommand{\baselinestretch}{0.97}
\caption{Integrated intensity map of the Orion Molecular Ridge in \nwater\ 1$_{10}$-1$_{01}$ 
556.9~GHz transition using {\em SWAS}
(see Table~1).  The inner region, denoted by the bolded square encompassing the spectra centered
on position (0, 0) in the right panel, includes
gas also subject to strong outflow shocks and is not included in the analysis.  The starred numbers
within this square indicate the values by which the antenna temperatures  have been divided 
so that the scaled spectra fit within this plot.
All offsets are relative to $\alpha\,=\,$05$^{\rm h}$35$^{\rm m}$14$^{\rm s}\!\!.$5,
$\delta\:=\,-$05\ddeg 22\amin 37\asec\ (J2000). }
\renewcommand{\baselinestretch}{1.0}
\label{fig:map3}
\end{figure}

\clearpage

\begin{figure}[t]
\centering
\vspace{-6mm}
\includegraphics[scale=0.82]{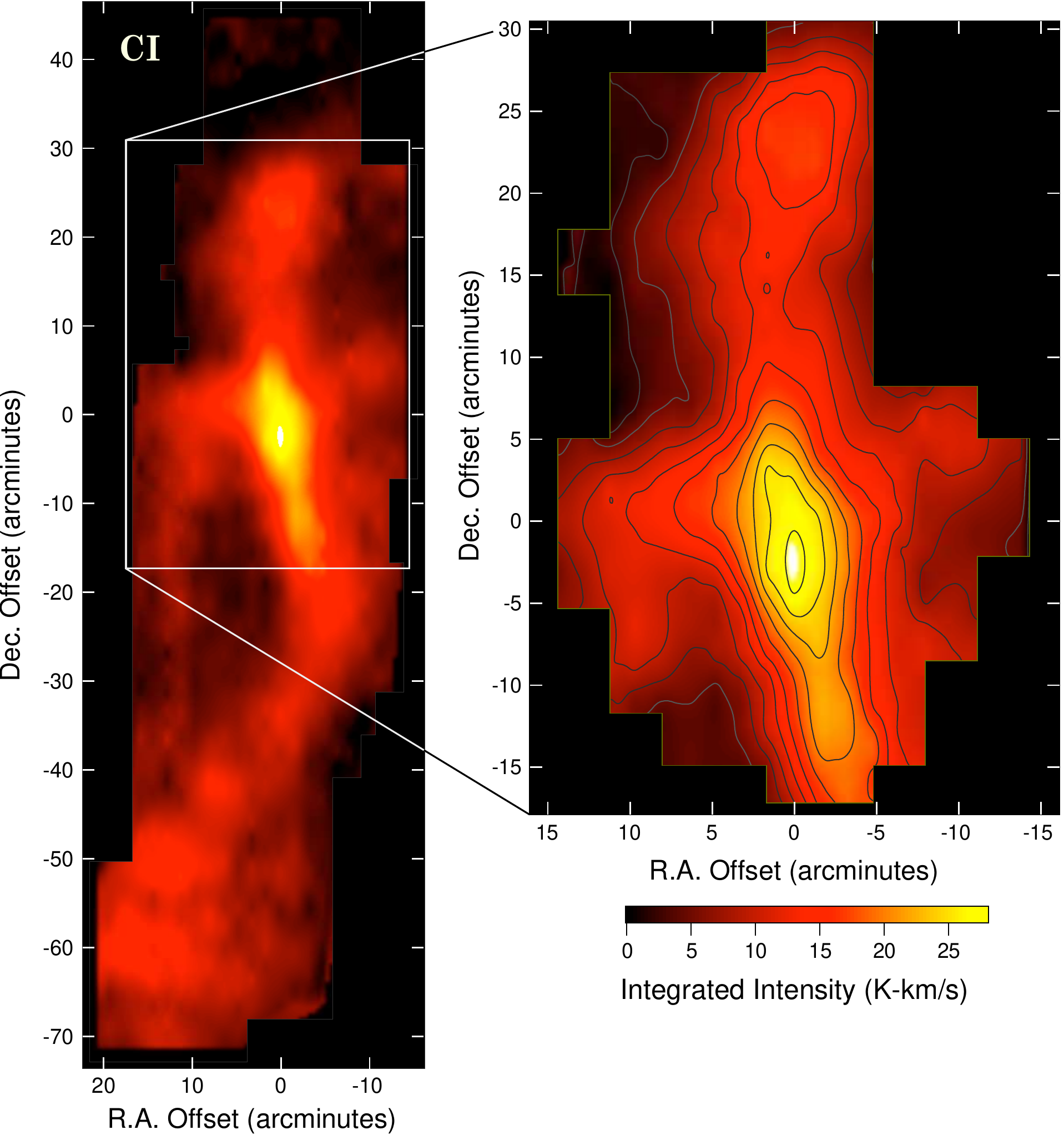}
\vspace{0.1mm}
\renewcommand{\baselinestretch}{0.97}
\caption{{\em Left:}~\swas\ integrated intensity map of the Orion Molecular Ridge in the 
CI $^3$P$_1-^3$P$_0$
492.2~GHz transition (see Table~1).  {\em Right:}~Subset of the extended \swas\
CI map obtained with the longer integration times used to measure the \nwater\ 1$_{10}$-1$_{01}$
556.9~GHz emission.  The peak CI integrated intensity is 28.3~K km s$^{-1}$
and the contours are in increments of 2.5~K km s$^{-1}$ from zero integrated intensity. 
All offsets are relative to $\alpha\,=\,$05$^{\rm h}$35$^{\rm m}$14$^{\rm s}\!\!.$5,
$\delta\:=\,-$05\ddeg 22\amin 37\asec\ (J2000). }
\renewcommand{\baselinestretch}{1.0}
\label{fig:map4}
\end{figure}

\clearpage

\begin{figure}[t]
\centering
\vspace{-6mm}
\includegraphics[scale=0.82]{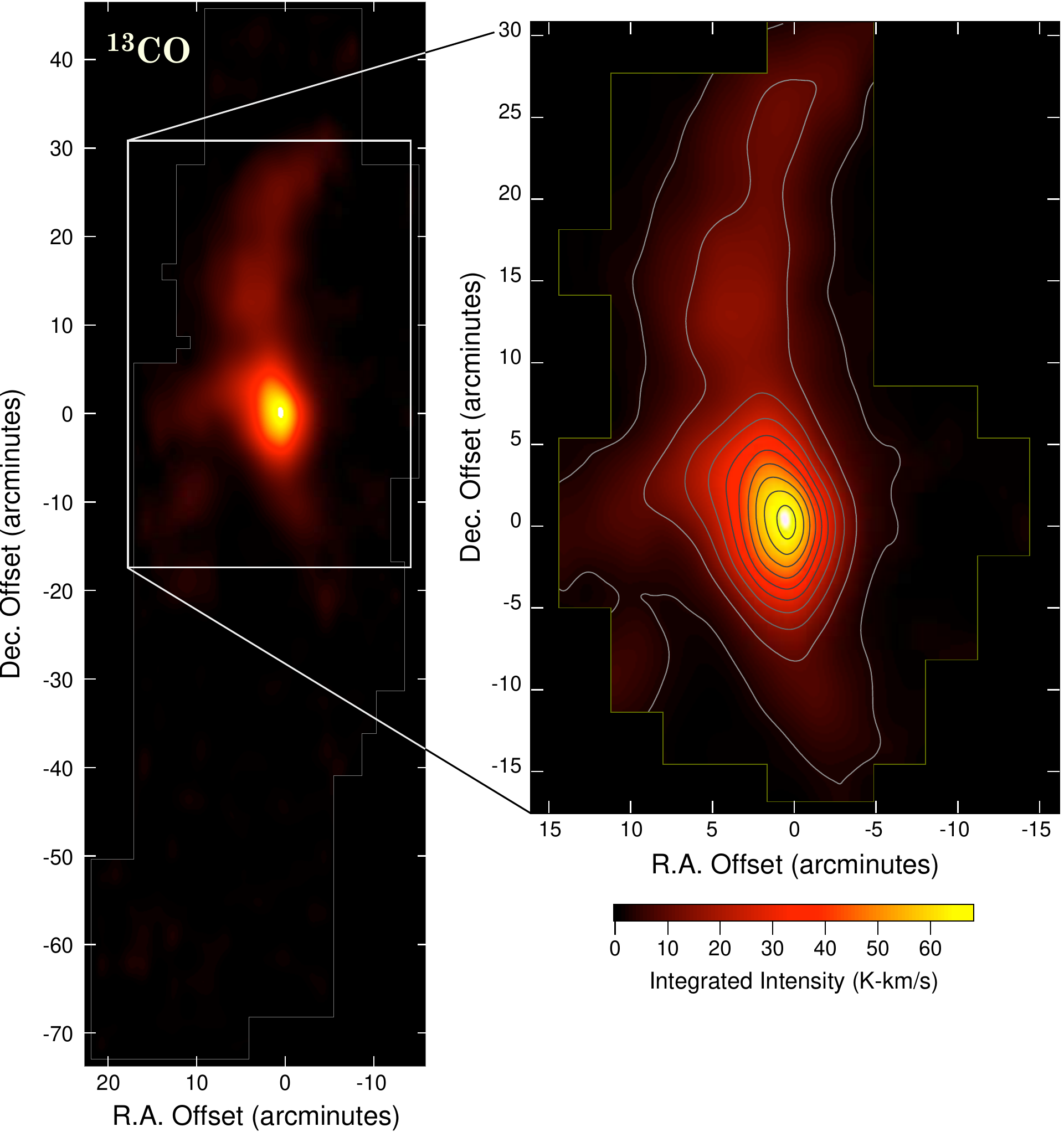}
\vspace{0.1mm}
\renewcommand{\baselinestretch}{0.97}
\caption{{\em Left:}~\swas\ integrated intensity map of the Orion Molecular Ridge in the 
$^{13}$CO $J=\,$5\dash 4
550.9~GHz transition (see Table~1).  {\em Right:}~Subset of the extended \swas\
$^{13}$CO $J=\,$5\dash 4 map obtained with the longer integration times used to measure
the \nwater\ 1$_{10}$-1$_{01}$
556.9~GHz emission.  The peak $^{13}$CO $J=\,$5$\,$--$\,$4 integrated intensity is 
69.1~K km s$^{-1}$ and the contours are in increments of 7~K km s$^{-1}$ from an 
integrated intensity of 3~K km s$^{-1}$. 
All offsets are relative to $\alpha\,=\,$05$^{\rm h}$35$^{\rm m}$14$^{\rm s}\!\!.$5,
$\delta\:=\,-$05\ddeg 22\amin 37\asec\ (J2000). }
\renewcommand{\baselinestretch}{1.0}
\label{fig:map5}
\end{figure}

\clearpage

\vspace{0.6in}

\begin{figure}[t]
\centering
\vspace{-5.0mm}
\includegraphics[scale=0.72]{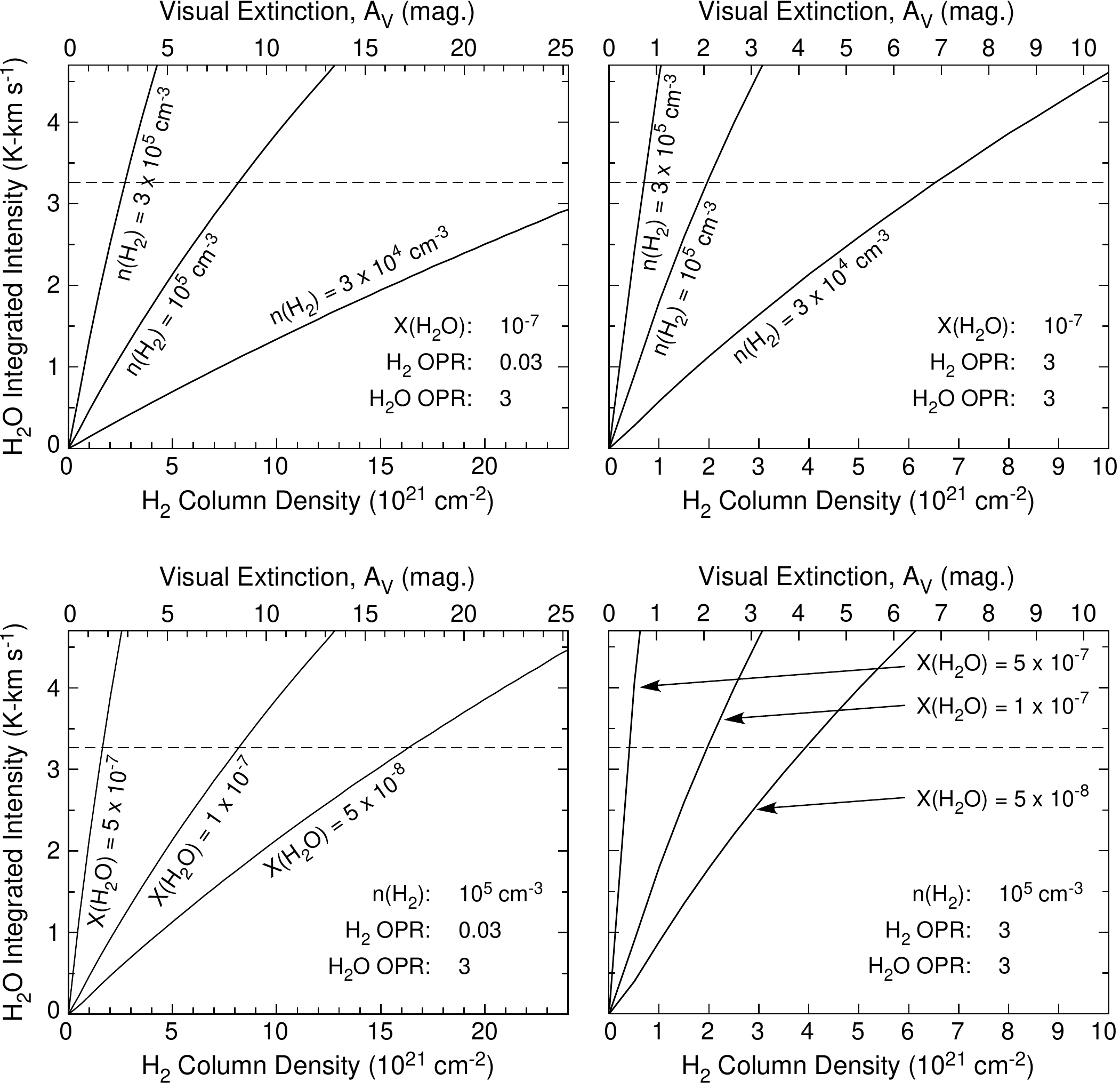}
\renewcommand{\baselinestretch}{0.998}
\vspace{1.5mm}
\caption{Plots of computed \water\ 1$_{10}-$1$_{01}$ integrated
antenna temperature for a Gaussian line, 1.064$\,\times$$\int T_A dv$, versus \mh\ column density.
The calculations assume a slab geometry, a gas temperature of 30~K,
and a line width of 3.5~km s$^{-1}$.
The horizontal dashed line in each plot denotes the maximum integrated
antenna temperature measured among the 77 spatial positions sampled, 
3.26 K km~s$^{-1}$.  Thus, the \water\ integrated intensities
from all observed positions can be reproduced by conditions below the
dashed line.  
{\em Upper left}: Curves of $\int T_A dv$ vs.~$N$(H$_2$) for
$X$(\water)$\:=\:$10$^{-7}$, an \mh\ OPR$\:=\:$0.03, the LTE value at $T\:=\:$30~K,
an \water\ OPR$\:=\:$3, and \mh\ densities of 3$\times$10$^4$, 10$^5$, and 3$\times$10$^5$
\cmc.  {\em Upper right}: Same as the {\em upper left} plot, except the \mh\ OPR is assumed to
be 3.  {\em Lower left}:  Curves of $\int T_A dv$ vs.~$N$(H$_2$) for
$n$(H$_2$)$\:=\:$10$^5$, an \mh\ OPR$\:=\:$0.03, an \water\ OPR$\:=\:$3,
and assumed total (ortho$+$para) \water\ abundances
of 5$\times$10$^{-8}$, 10$^{-7}$, and 5$\times$10$^{-7}$.
{\em Lower right}: Same as the {\em lower left} plot, except the \mh\ OPR is assumed to
be 3. The depth into the cloud, measured in magnitudes of visual extinction, assumes
$N$(H$_2$)$\:=\:$9.5$\times$10$^{20}\,$\av\ \cms.}
\renewcommand{\baselinestretch}{1.0}
\label{effectivelythin}
\end{figure}

\clearpage

\begin{figure}[t]
\centering
\vspace{-0.2in}
\includegraphics[scale=0.70]{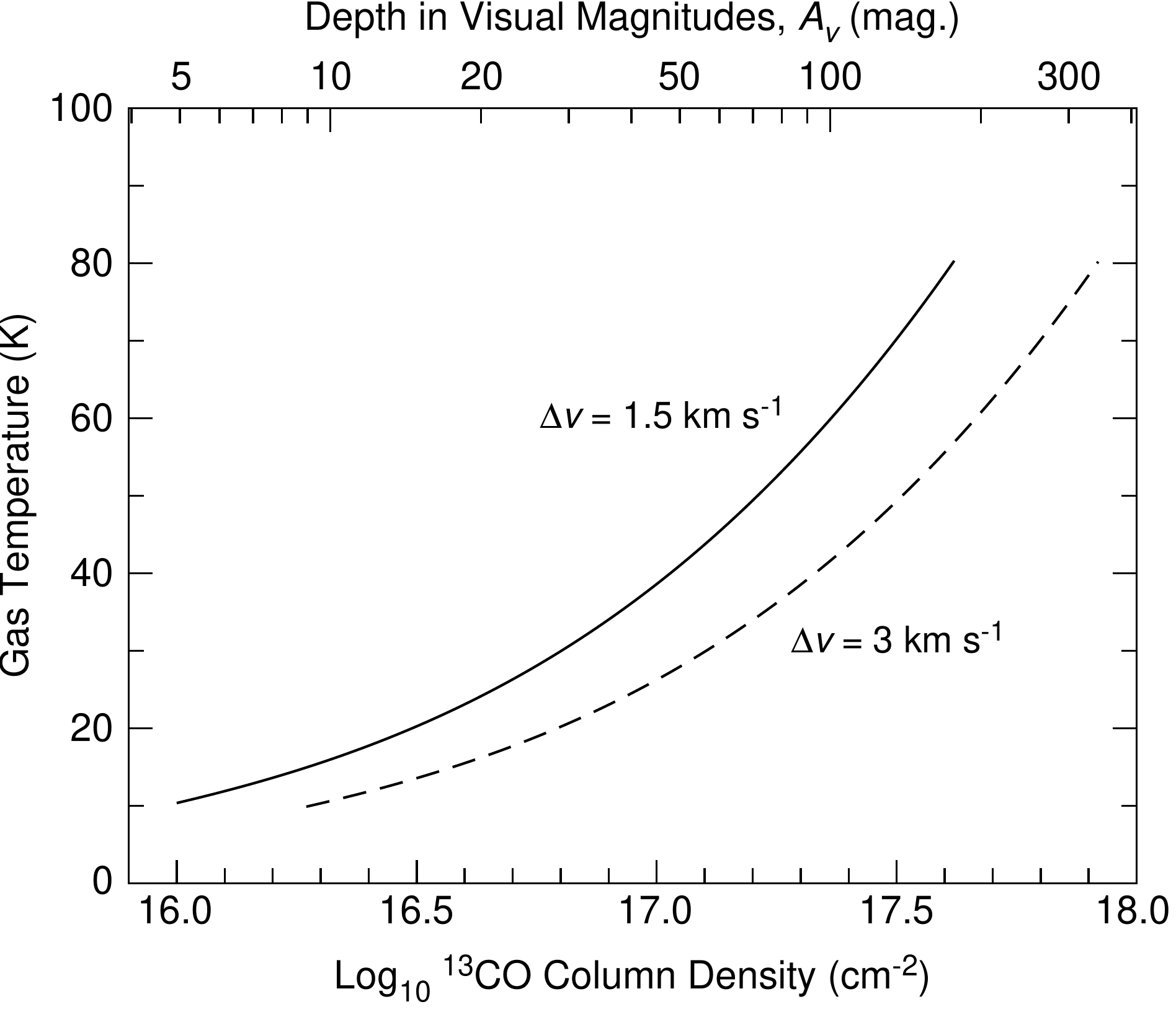}
\vspace{6mm}
\renewcommand{\baselinestretch}{0.98}
\caption{Plot of the column density required to reach an optical depth of 1 in the 
\ico\ $J\,=\:$1$\,$--$\,$0 transition as a function of the gas temperature for line
widths of 1.5 and 3\kms\ (see text).  The relation between the \ico\ column density and the depth
into the cloud, measured in 
magnitudes of visual extinction, is that provided in Pineda~\etal\ (2008).} 
\renewcommand{\baselinestretch}{1.0}
\label{13cotau1}
\end{figure}

\clearpage

\begin{figure}[t]
\centering
\vspace{-6mm}
\includegraphics[scale=0.72]{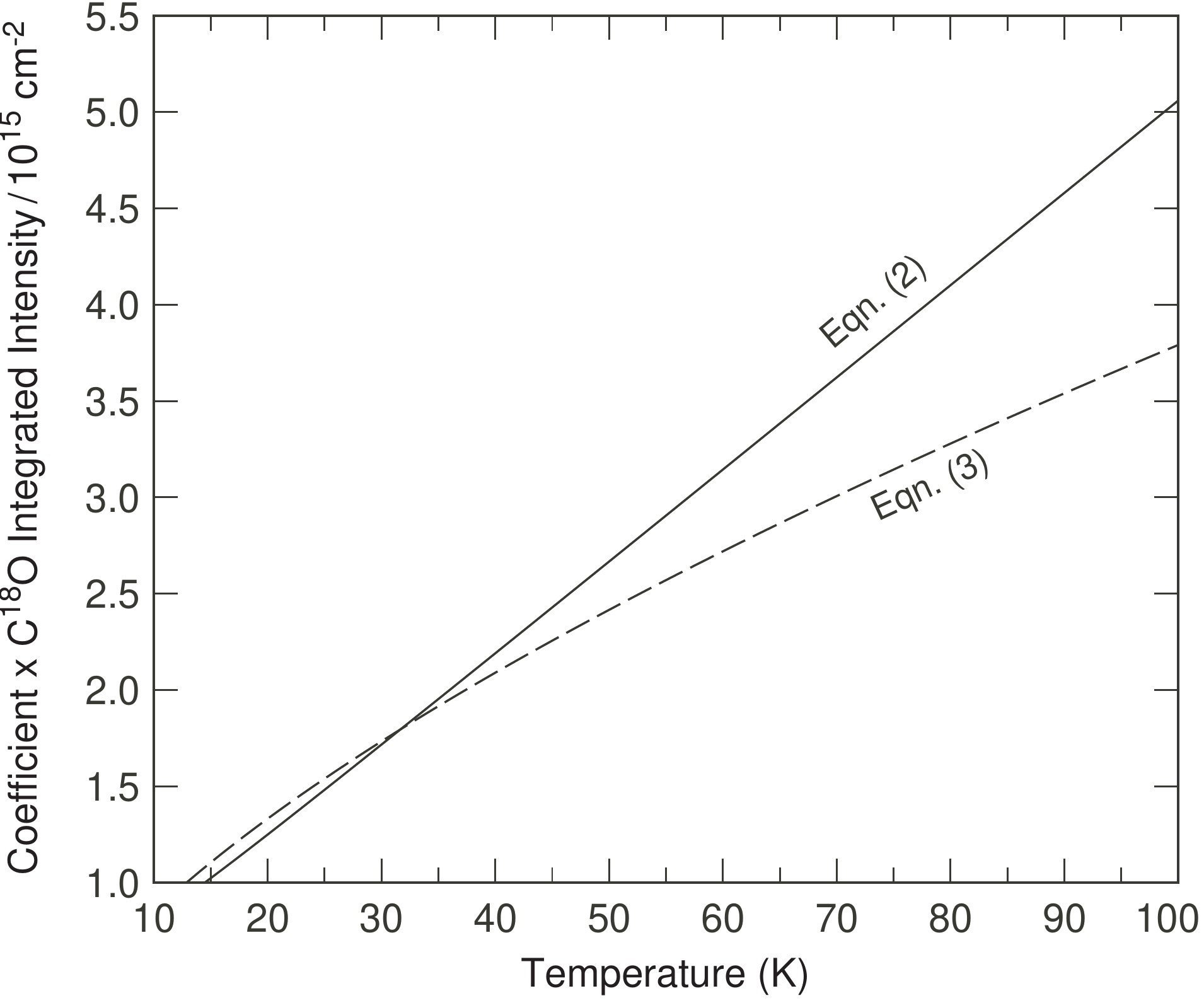}
\vspace{8.0mm}
\renewcommand{\baselinestretch}{0.98}
\caption{Comparison of the leading terms in Eqns.~(2) and (3), which provide two 
different approaches to connecting the \cio\ column density, $N$(C$^{18}$O), to the 
integrated intensity of the \cio\ $J =\:$1\dash 0 line.}
\renewcommand{\baselinestretch}{1.0}
\label{coefplot}
\end{figure}

\clearpage

\begin{figure}[t]
\vspace{-1.0in}
\centering
\includegraphics[scale=0.72]{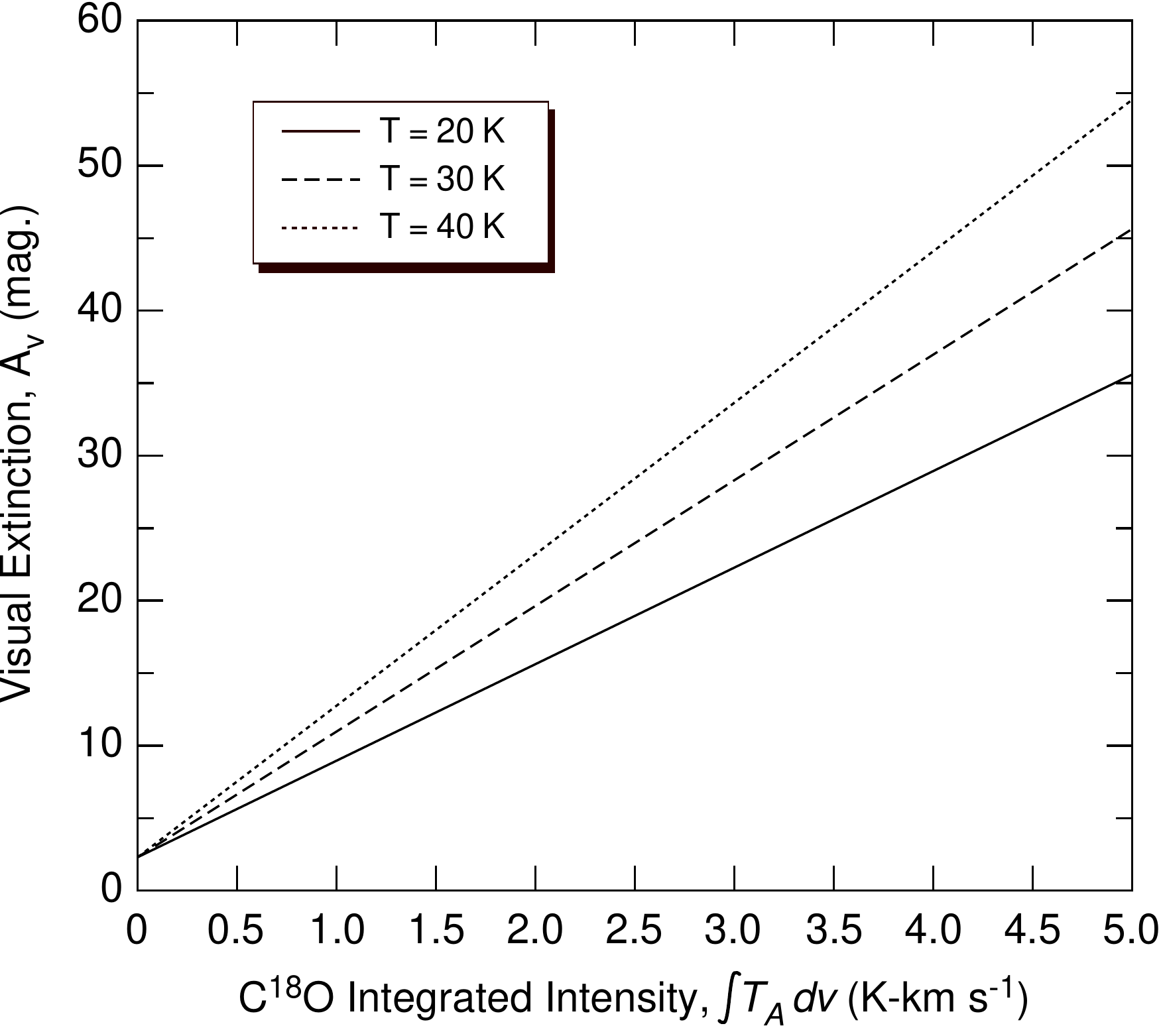}
\vspace{9mm}
\caption{Plot of temperature-dependent relation between the visual extinction, A$_V$, 
and the C$^{18}$O integrated intensity, $\int T_A dv$ (K km s$^{-1}$), given in Eqn.~(4),
corrected for the main-beam efficiency, 0.5, at the C$^{18}$O frequency.
A temperature of 30~K is assumed in our analysis.  The measured \cio\ integrated
intensities toward the positions considered in this study are all less than 4 K km s$^{-1}$.}
\label{avplot}
\end{figure}

\clearpage

\begin{figure}[t]
\centering
\vspace{-6mm}
%$\!$\includegraphics[scale=0.91]{fig11.eps}
$\!$\includegraphics[scale=0.91]{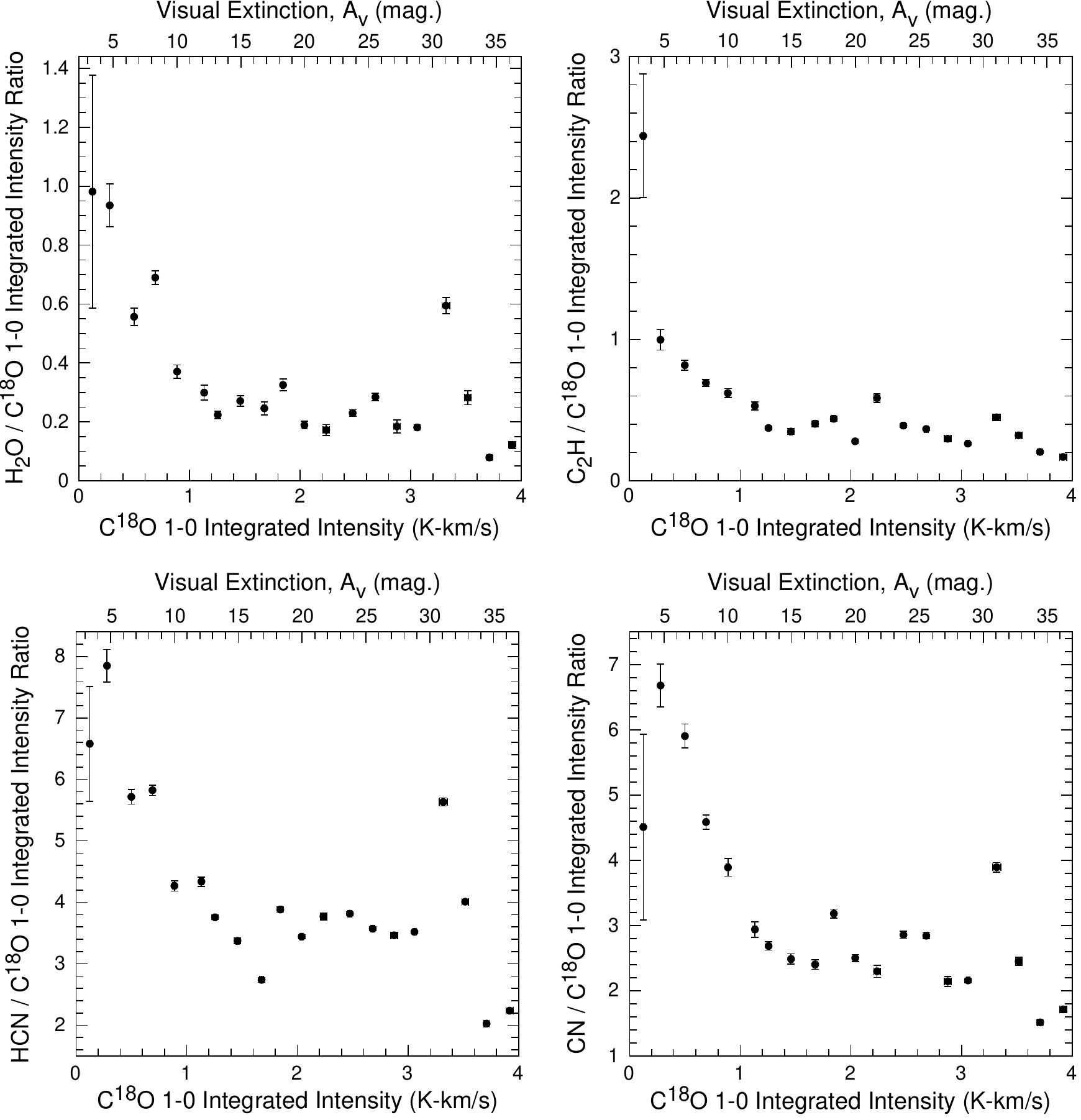}
%$\!$\includegraphics[scale=0.91]{Ratio_fig1.pdf}
\vspace{1mm}
\renewcommand{\baselinestretch}{0.97}
\caption{Plots of the ratio of the H$_2$O, C$_2$H, HCN, and CN integrated intensities to 
the \cio\ $J=\,$1\dash 0 integrated intensity versus the \cio\ $J=\,$1\dash 0 integrated intensity.  
Using Eqn.~(4) and
assuming a gas temperature of 30$\:$K, the ratios are also presented as a function of
visual magnitude, \av.  The 77 spatial positions observed have been binned according
to their \cio\ integrated intensities and co-averaged in 20 equal 
$x$-axis bins of 0.2 K km s$^{-1}$ to reduce the
dispersion in each plot.  The $x$-$y$ error bars for each point represent the error-weighted
mean and 1$\sigma$ uncertainty in the mean for the co-averaged points in each bin.
The high point at a \cio\ integrated intensity of 3.3 K km s$^{-1}$ represents one spatial
sample corresponding to position $\Delta\alpha, \Delta\delta\:=\:$6.4, 0.  The proximity of
this one sight line to BN/KL is likely responsible for the elevated ratio values seen in these plots.}
\renewcommand{\baselinestretch}{1.0}
\label{Ratio_fig1}
\end{figure}

\clearpage

\begin{figure}[t]
\centering
\vspace{-6mm}
%$\!\!$\includegraphics[scale=0.91]{fig12.eps}
$\!\!$\includegraphics[scale=0.91]{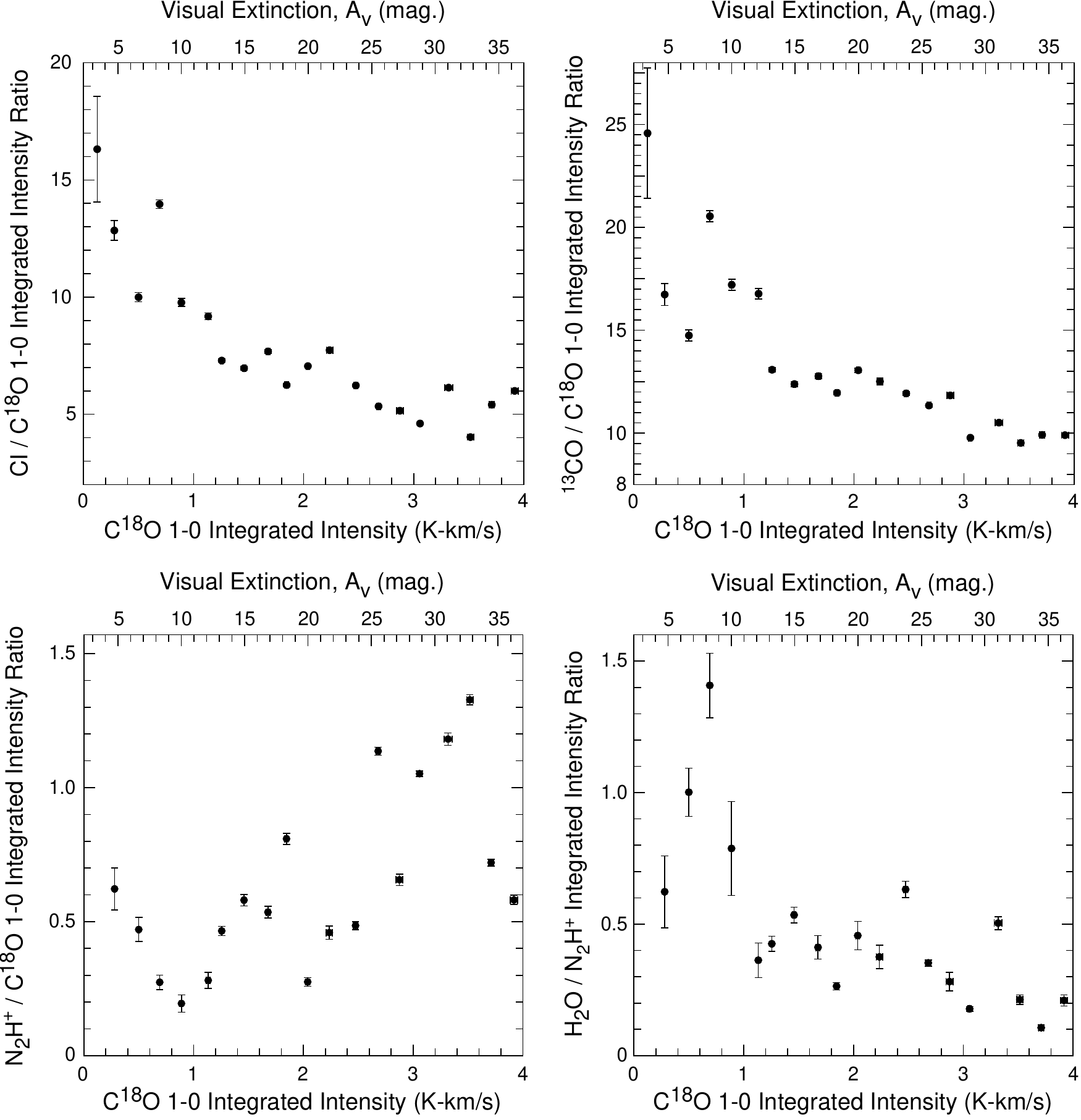}
%$\!\!$\includegraphics[scale=0.91]{Ratio_fig2.pdf}
\vspace{2mm}
\renewcommand{\baselinestretch}{0.97}
\caption{Same as Fig.~11, except for the ratio of the CI, \ico, and N$_2$H$^+$ 
integrated intensities to the \cio\ $J=\,$1-0 integrated intensity and the H$_2$O to N$_2$H$^+$
integrated intensities to the \cio\ $J=\,$1-0 integrated intensity.  Note that one point
at the lowest \cio\ integrated intensity is excluded from the bottom two panels due to the
absence of detectable N$_2$H$^+$ emission from these positions.}
\renewcommand{\baselinestretch}{1.0}
\label{Ratio_fig2}
\end{figure}

\clearpage

\begin{figure}[t]
\centering
\vspace{-6mm}
\includegraphics[scale=0.775]{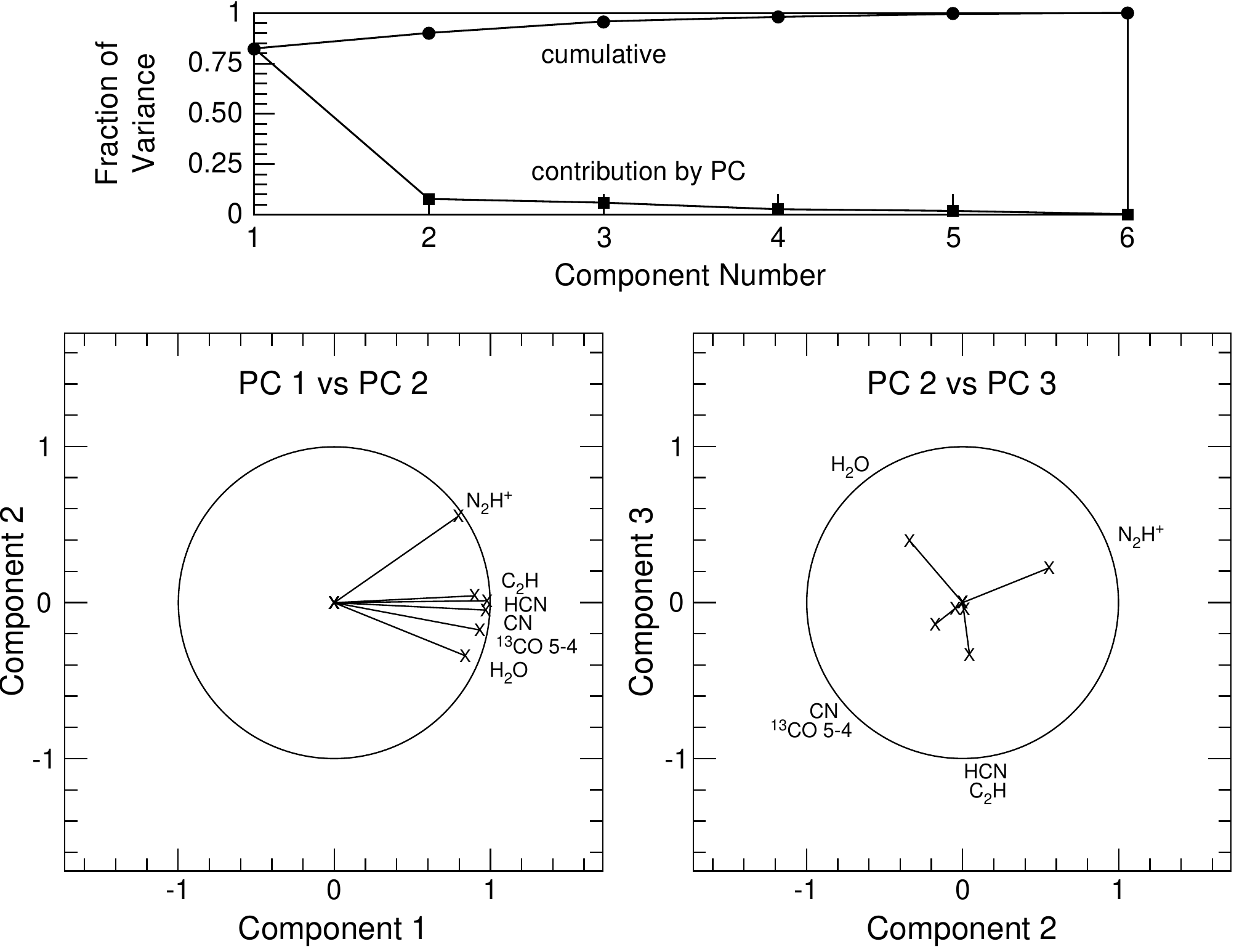}
\vspace{5.1mm}
\renewcommand{\baselinestretch}{0.97}
\caption{Results of PCA analysis for high (i.e., $>\,$10$^4$\cmc)
critical density species. {\em Top}: Fraction of the total variance accounted for by each 
principal component, along with the cumulative fraction for the first $n$ principal
components as a function of $n$.  {\em Bottom left}: Coefficients for the first 
and second principal components needed to approximate the maps of each transition. 
{\em Bottom right}: Coefficients for the second and third principal components.}
\renewcommand{\baselinestretch}{1.0}
\label{pca}
\end{figure}

\clearpage

\begin{figure}[t]
\centering
\vspace{-0.2in}
\includegraphics[scale=0.70]{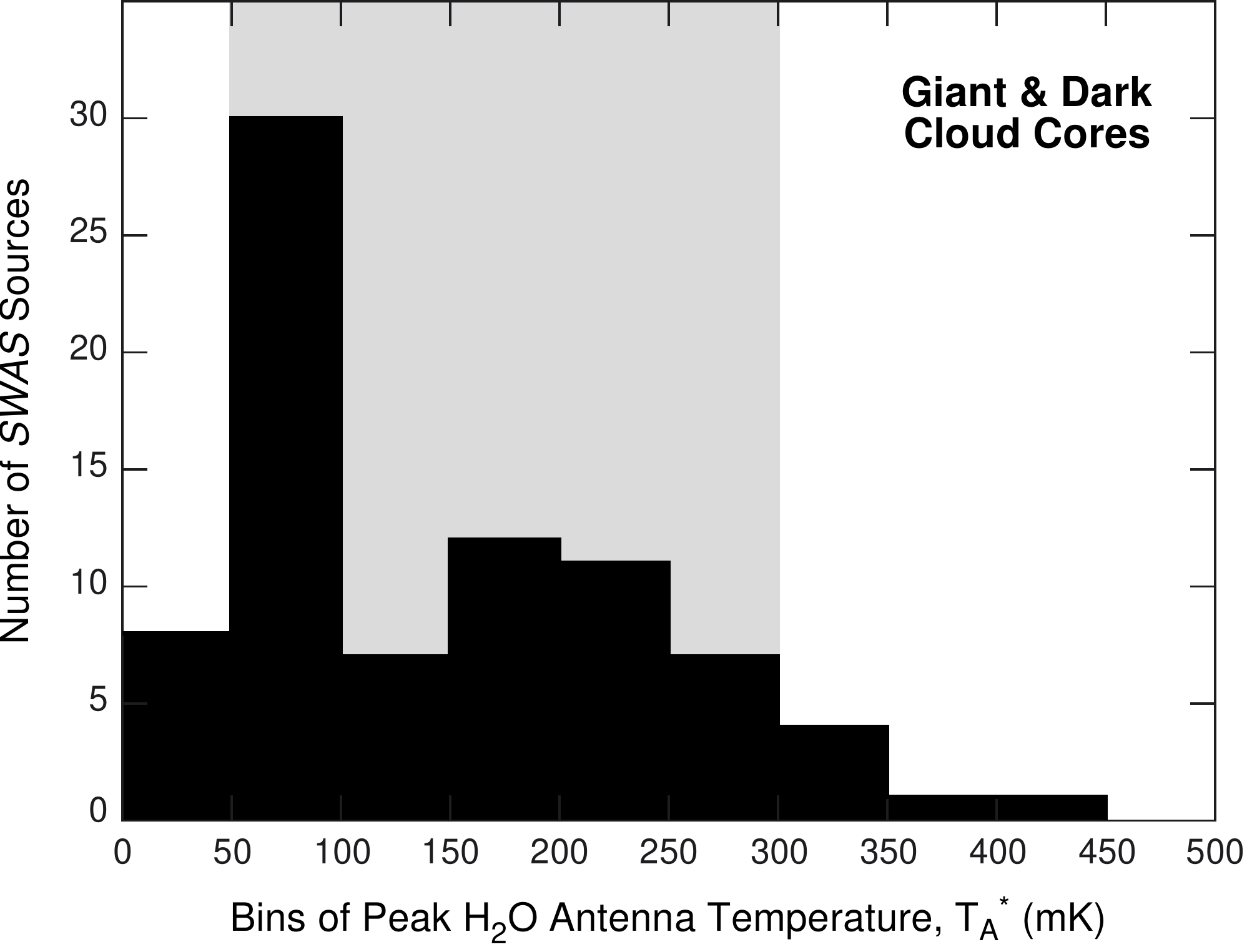}
\vspace{11mm}
\renewcommand{\baselinestretch}{0.97}
\caption{Histogram of peak H$_2^{\:16}$O 1$_{10}-$1$_{01}$ 556.9~GHz antenna temperatures,
\ta, measured toward 83 dark and giant cloud cores by {\em SWAS}. Almost 70 percent 
of the sources were observed to have peak \ta's within a factor of two of 100 mK, while 
more than 80 percent of the sources exhibit peak \ta's between 50 and 300 mK (grey area).  This distribution is much narrower than would be expected based on the spread 
of column densities and volume densities of the sources in the sample (see text).}
\renewcommand{\baselinestretch}{1.0}
\label{histogram}
\end{figure}

\clearpage

\begin{figure}[t]
\centering
\vspace{0.60in}
\includegraphics[scale=0.75]{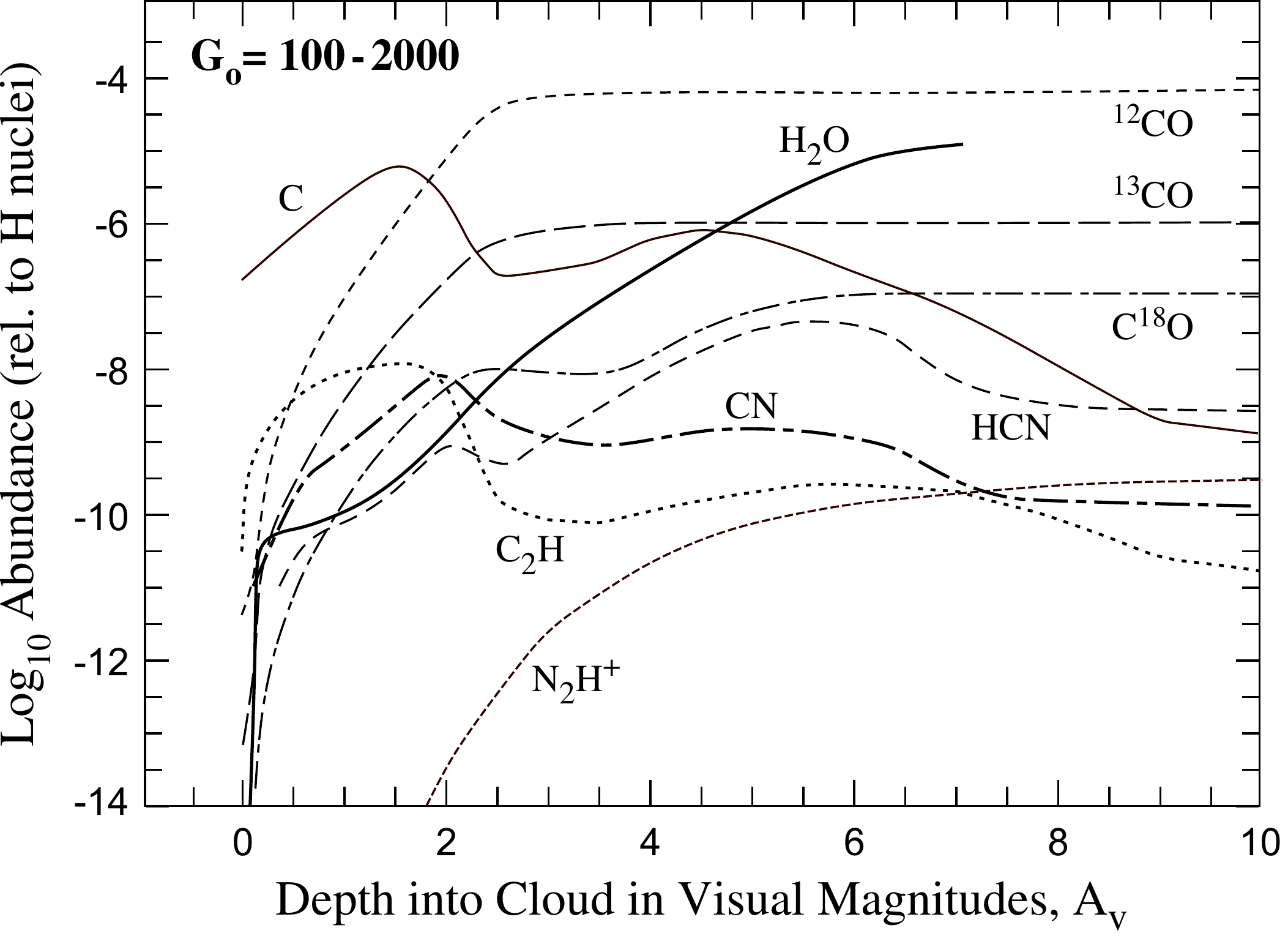}
\renewcommand{\baselinestretch}{0.97}
\vspace{6.5mm}
\caption{Predicted abundance profiles for the densities and moderate incident far-ultraviolet field
strengths applicable to the extended Orion ridge.  Specifically, the CN and HCN profiles
are adapted from Boger \& Sternberg (2005) who assumed a hydrogen 
nuclei density, $n_H$, of 10$^4$~\cmc\ and a FUV intensity, $G_{\rm o}$, of $2 \times 10^3$.
The C, C$_2$H and N$_2$H$^+$ profiles are adapted from Morata \& Herbst (2008)
who assumed $n_H =\:$2$\times$10$^4$~\cmc\ and $G_{\rm o} =\:$100.  
The $^{12}$CO, \ico, \cio\ and \water\ profiles were
adapted from Jansen~\etal\ (1995a) who assumed 
$n_H =\:$10$^5$~\cmc\ and $G_{\rm o} =\:$650.}
\renewcommand{\baselinestretch}{1.0}
\label{abundanceprofile-lowG}
\end{figure}

\clearpage

\begin{figure}[t]
\centering
\vspace{0.60in}
\includegraphics[scale=0.75]{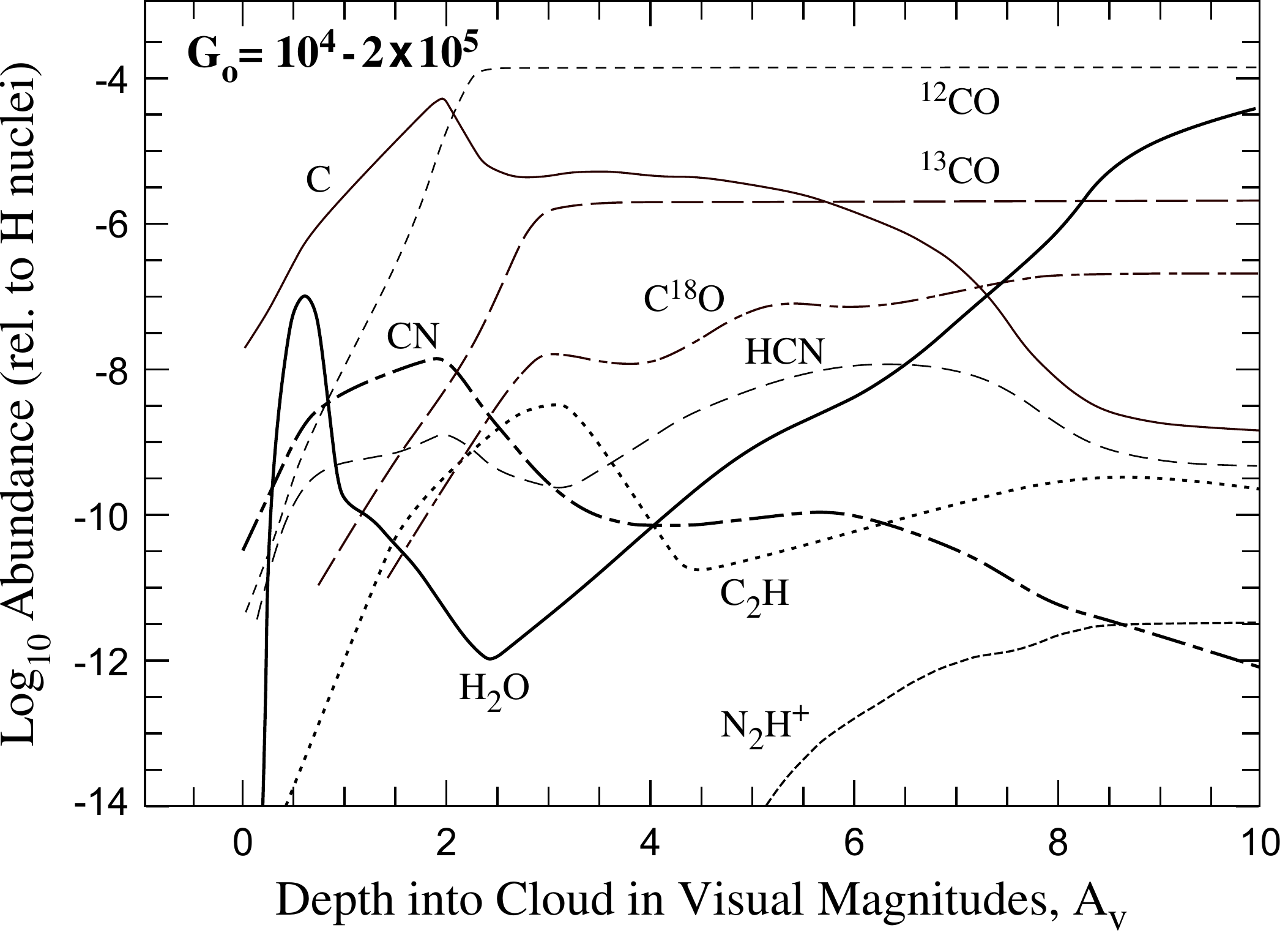}
\renewcommand{\baselinestretch}{0.97}
\vspace{6.5mm}
\caption{Predicted abundance profiles for the densities and strong incident far-ultraviolet field
strengths applicable to the Orion ridge close to the Trapezium cluster (see text).  
Specifically, the C, $^{12}$CO, CN, and HCN profiles
are adapted from Boger \& Sternberg (2005) who assumed a hydrogen 
nuclei density, $n_H$, of 10$^5$~\cmc, and a FUV intensity, $G_{\rm o}$, of $2 \times 10^4$.
The C$_2$H profile is adapted from Morata \& Herbst (2008)
who assumed $n_H =\:$2$\times$10$^4$~\cmc\ and $G_{\rm o} =\:$10$^4$.  
The \ico\ and \cio\ profiles were
adapted from Jansen~\etal\ (1995b) who assumed 
$n_H =\:$2.5$\times$10$^5$~\cmc\ and $G_{\rm o} =\:$4.4$\times$10$^4$.
The \water\ and N$_2$H$^+$ profiles are adapted from Sternberg \& Dalgarno (1995)
who assumed $n_H =\:$10$^6$~\cmc\ and $G_{\rm o} =\:$2$\times$10$^5$.}
\renewcommand{\baselinestretch}{1.0}
\label{abundanceprofile-highG}
\end{figure}

\clearpage

\begin{figure}[t]
\centering
\vspace{-0.30in}
%$\!$\includegraphics[scale=0.725]{fig17.eps}
$\!$\includegraphics[scale=0.725]{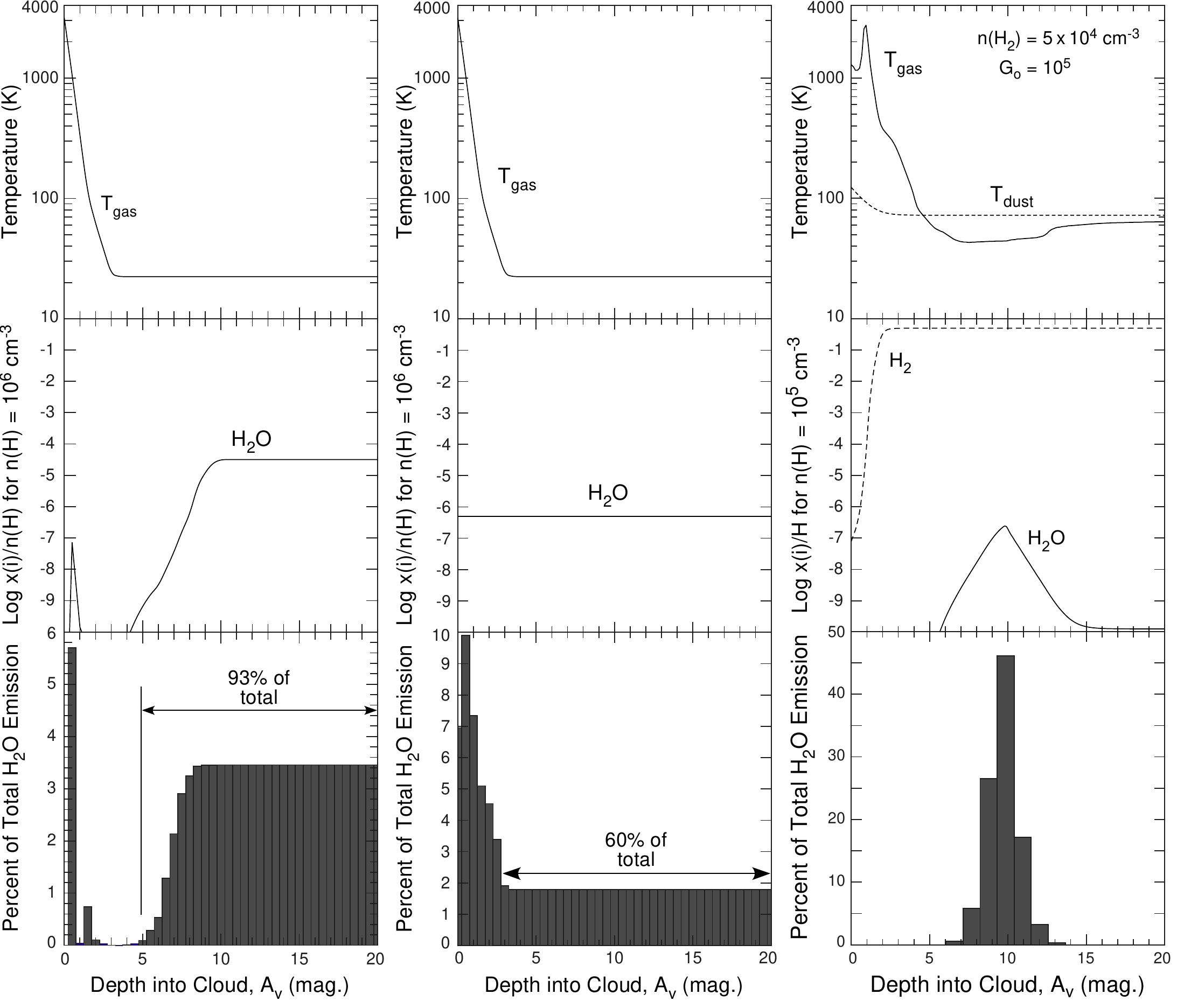}
%$\!$\includegraphics[scale=0.725]{H2O_Excitation.pdf}
\renewcommand{\baselinestretch}{0.97}
\vspace{-5mm}
\caption{Calculated fraction of the total \water\ emission arising from various depths 
into a dense cloud between \av$=\:$0 to 20.  {\em Left panel}: Profiles
of temperature and \water\ abundance obtained from
Sternberg \& Dalgarno (1995) for a PDR with
\nh$\:=\:$5$\times$10$^5$~\cmc\ and $G_{\rm o} =\:$2$\times$10$^5$.  The profiles in their paper 
cover the range 0$\,<\,$\av$\,<\,$10; our calculations were extended to an \av\ of 20 
by assuming that both the temperature and \water\ abundance at \av$ = $10 have 
reached equilibrium values that apply between \av$ =\:$10 and 20.
{\em Middle panel}: For the case of a constant water abundance, we use the Sternberg and
Dalgarno PDR density and temperature profile, but assume a constant \water\ abundance of
5$\times$10$^{-7}$, consistent with maximum \water\ abundance predicted by 
Hollenbach~\etal\ (2009).  {\em Right panel}:  Profiles
of temperature and \water\ abundance from Hollenbach~\etal\ (2009) model.  
In all cases, the collisional rates
of Faure~\etal\ (2007) were used and ortho-to-para \mh\ and 
\water\ ratios of 3:1 and total line width of 3~\kms\ were assumed.}
\renewcommand{\baselinestretch}{1.0}
\label{h2oexcitation}
\end{figure}

\clearpage

\begin{figure}[t]
\centering
\vspace{-0.90in}
%$\!$\includegraphics[scale=0.605]{fig18.eps}
$\!$\includegraphics[scale=0.605]{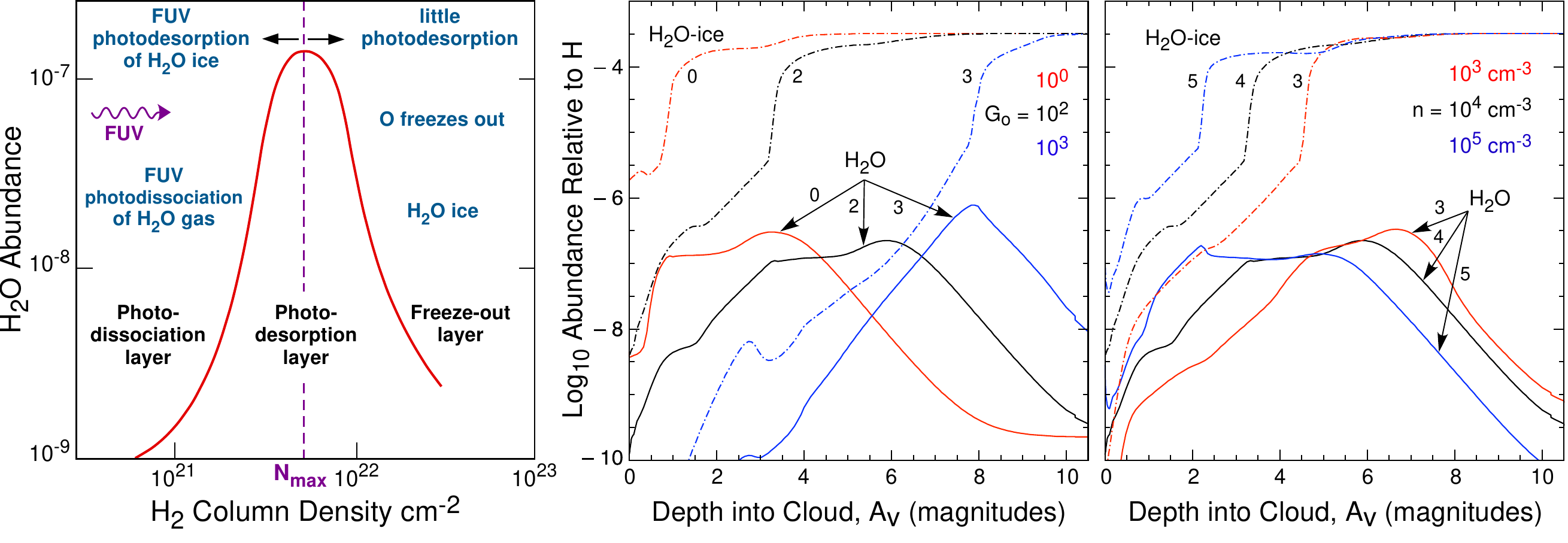}
%$\!$\includegraphics[scale=0.605]{Hollenbachfig.pdf}
\renewcommand{\baselinestretch}{0.97}
\vspace{-1mm}
\caption{{\em Left}:~Schematic depiction of water-vapor zone within molecular clouds,
including the main processes affecting the gas-phase water abundance.  {\em Center}:~\water\
and \water-ice abundances for a cloud with $n = $10$^4$ cm$^{-3}$ but with a variety of FUV
field strengths incident on the cloud surface (after Hollenbach~\etal\ 2009).
{\em Right}:~Effect of changing the gas density 
for the FUV field strength \go\ = 10$^2$ where,
following the convention used in Hollenbach~\etal\ (2009),
$n$ is the gas-phase 
hydrogen nucleus number density [$\sim n$(H) + 2$n$(\mh\ + $n$(H$^+$)].
The Hollenbach~\etal\ (2009) results depict
steady state abundance profiles, including CO depletion due to reactions with He$^+$.
Prior to the depletion of CO, the gas-phase \water\ abundances at high \av\ will be greater than
shown here, though the peak of water abundance at intermediate \av\ is likely retained.
(see text).}
\renewcommand{\baselinestretch}{1.0}
\label{schematicprofile}
\end{figure}

\pagebreak

\begin{table}[t]
\begin{center}
\label{tab:1}
\rule{0mm}{8mm}TABLE~1.\\*[1.5mm]
Spectral Lines Observed by {\em SWAS} and {\em FCRAO}\\*[1.6mm]
\begin{tabular}{ccccc} \hline \\*[-5.3mm] \hline
\multicolumn{5}{c}{{\rule{0mm}{4.9mm} \em SWAS}} \\*[0.5mm] \hline
\rule{0mm}{4.4mm}  &  & ~~Energy Above~~ &   & Critical \\*[-0.9mm]
~~Species~~ & ~~Transition~~ & Ground State &
  ~~~Frequency~~~ &  ~~~~~~Density~~~~~~ \\*[-0.3mm]
  &  &  ({\em E$_u$/k}) & (GHz) & (\cmc) \\*[0.5mm] \hline
\rule{0mm}{5.0mm}O$_2$ & 3,3$\,-\,$1,2 & 26~K & 487.249 & 
  10$^3$ \\*[0.70mm]
CI  & \phantom{$^{\, a}$}$^3$P$_1\,-\,^3$P$_0$$^{\, a}$ & 24~K & 492.161 &
  10$^3$ \\*[0.70mm]
H$_2^{\:18}$O & \phantom{$^{\, a}$}1$_{10}\,-\,$1$_{01}$$^{\, a}$ & 26~K & 547.676 &
  \phantom{$\;^{\, b}$}8$\times$10$^{7}\,^{\, b}$ \\*[0.70mm]
\ico\ & {\em J}$\:=\,$5 $-$ 4 & 79~K & 550.926 & 
  $\;$2$\times$10$^5$ \\*[0.70mm]
H$_2^{\:16}$O & \phantom{$^{\, a}$}1$_{10}\,-\,$1$_{01}$$^{\, a}$ & 27~K & 556.936 &
  \phantom{$\;^{\, b}$}8$\times$10$^{7}\,^{\, b}$\\*[0.7mm] \hline
\multicolumn{5}{c}{{\rule{0mm}{4.9mm} \em FCRAO}} \\*[0.5mm] \hline
\rule{0mm}{5.0mm}C$_2$H & ~~N=1$-$0, J=$\frac{1}{2}-\frac{1}{2}$~~ &
  4.20~K & \phantom{1$^{\, c}$}87.402$^{\, c}$ & 2$\times$10$^{5}$ \\*[0.70mm]
HCN & J=1$-$0 & 4.25~K & \phantom{1$^{\, c}$}88.632$^{\, c}$ &  2$\times$10$^6$ \\*[0.70mm]
N$_2$H$^+$ & J=1$-$0 & 4.47~K & \phantom{1$^{\, c}$}93.174$^{\, c}$ & 
2$\times$10$^5$ \\*[0.70mm]
C$^{18}$O & J=1$-$0 & 5.27~K & 109.782 & 2$\times$10$^3$ \\*[0.70mm]
$^{13}$CO & J=1$-$0 & 5.29~K & 110.201 & 2$\times$10$^3$   \\*[0.70mm]
CN & N=1$-$0, J=$\frac{3}{2}-\frac{1}{2}$ &
5.45~K & \phantom{$^{\, c}$}113.491$^{\, c}$ & 4$\times$10$^4$ \\*[0.70mm] 
$^{12}$CO & J=1$-$0 & 5.53~K & 115.271 &  2$\times$10$^3$  \\*[0.70mm] \hline
\end{tabular}
\end{center}
\vspace{-6.0mm}
\begin{list}{}{\leftmargin 0.32in \rightmargin 0.24in \itemindent -0.11in 
 \baselineskip 14.0pt}
\item \raisebox{0.20ex}{$^{a}$}~Ground-state transition.\\*[-7.2mm]
\item \raisebox{0.20ex}{$^{b}$}~Based on the collisional rates of 
Dubernet~\etal\ (2006) and
Faure~\etal\ (2007) and assuming collisions with 
ortho- and para-\mh\ in the ratio of 0.03, the LTE value at 30~K.  The critical density for 
H$_2^{\:16}$O will likely be less than this value due to significant radiation 
trapping in this line.  The critical density for H$_2^{\:18}$O could be reduced 
due to the same effect (see text).\\*[-7.2mm]
\item \raisebox{0.20ex}{$^{c}$}~Rest frequency of the strongest hyperfine component.
\end{list}
\end{table}

\end{document}